\documentclass[arguments]{aastex631}

\usepackage{amsmath}
\usepackage{microtype}
\usepackage{multirow}

\begin{document}

\title{Measuring the redshift-space distortions by cross-correlating the density fields before and after reconstruction}

\author[0000-0001-6613-2709]{Weibing Zhang}

\affiliation{National Astronomical Observatories, Chinese Academy of Sciences, Beijing, 100101, P.R.China}

\affiliation{School of Astronomy and Space Sciences, University of Chinese Academy of Sciences, Beijing, 100049, P.R.China}

\author{Ruiyang Zhao}

\affiliation{National Astronomical Observatories, Chinese Academy of Sciences, Beijing, 100101, P.R.China}

\affiliation{School of Astronomy and Space Sciences, University of Chinese Academy of Sciences, Beijing, 100049, P.R.China}

\affiliation{Institute of Cosmology and Gravitation, University of Portsmouth, Dennis Sciama Building, Portsmouth PO1 3FX, United Kingdom}

\author{Xiaoyong Mu}

\affiliation{National Astronomical Observatories, Chinese Academy of Sciences, Beijing, 100101, P.R.China}

\affiliation{School of Astronomy and Space Sciences, University of Chinese Academy of Sciences, Beijing, 100049, P.R.China}

\author{Kazuya Koyama}
\affiliation{Institute of Cosmology and Gravitation, University of Portsmouth, Dennis Sciama Building, Portsmouth PO1 3FX, United Kingdom}

\author{Ryuichi Takahashi}
\affiliation{Faculty of Science and Technology, Hirosaki University, 3 Bunkyo-cho, Hirosaki, Aomori 036-8561, Japan}

\author{Yuting Wang}
\affiliation{National Astronomical Observatories, Chinese Academy of Sciences, Beijing, 100101, P.R.China}

\affiliation{Institute for Frontiers in Astronomy and Astrophysics, Beijing Normal University, Beijing, 102206, P.R.China}

\correspondingauthor{Gong-Bo Zhao (\url{gbzhao@nao.cas.cn})}
\author[0000-0003-4726-6714]{Gong-Bo Zhao}
\affiliation{National Astronomical Observatories, Chinese Academy of Sciences, Beijing, 100101, P.R.China}

\affiliation{School of Astronomy and Space Sciences, University of Chinese Academy of Sciences, Beijing, 100049, P.R.China}

\affiliation{Institute for Frontiers in Astronomy and Astrophysics, Beijing Normal University, Beijing, 102206, P.R.China}

\begin{abstract}

In this work, we develop a theoretical model for the cross-power spectrum of the matter density field before and after standard baryonic acoustic oscillation (BAO) reconstruction. Using this model, we extract the redshift-space distortion (RSD) parameter from the cross-power spectrum. The model is validated against a suite of high-resolution \( N \)-body simulations, demonstrating its accuracy and robustness for cosmological analyses.

\end{abstract}

\keywords{}

\section{Introduction} 
\label{sec:intro}

The cosmic web, as mapped by large spectroscopic galaxy surveys, encodes rich cosmological information that can be extracted from the three-dimensional clustering patterns of galaxies \citep{Peebles:1980yev}. Clustering is generally quantified using $N$-point statistics, such as the $N$-point correlation function (NPCF) or the polyspectra of galaxies. Cosmological analyses traditionally focus on the two-point correlation function ($2$PCF) or the power spectrum, primarily because these observables are more straightforward to measure and interpret compared to higher-order statistics. In contrast, higher-order statistics are computationally expensive to calculate and require significantly larger sets of mock catalogs to accurately estimate the data covariance matrix, posing additional analytical challenges.

While two-point statistics capture crucial information about most cosmological parameters, higher-order statistics offer valuable complementary insights, particularly when small-scale modes are incorporated into the analysis. Significant progress has been made in leveraging higher-order statistics, including the development of efficient estimators for observables \citep{Philcox:2021bwo} and the construction of theoretical models for covariance matrices in cosmological analyses \citep{Philcox:2021hbm}.

Recently, a novel method was proposed to extract key components of higher-order statistics from two-point statistics \citep{Wang:2022nlx}. This approach leverages the BAO reconstruction technique, which effectively linearizes the galaxy density field. By jointly analyzing \( P^{\mathrm {pre}} \) (the power spectrum of the pre-reconstructed density field), \( P^{\mathrm {post}} \) (the power spectrum of the post-reconstructed density field), and \( P^{\mathrm{x}} \) (the cross-power spectrum between the pre- and post-reconstructed fields), \citet{Wang:2022nlx} demonstrated, via Fisher matrix analysis on Molino galaxy mock catalogs \citep{Molino} generated from the Quijote \( N \)-body simulations \citep{Quijote_sims}, that significant bispectrum information can be efficiently recovered. Applying this method to real galaxy catalogs requires theoretical models for \( P^{\mathrm{pre}} \), \( P^{\mathrm{post}} \), and \( P^{\mathrm{x}} \), which can be constructed using either emulator-based approaches \citep{prepostEmu} or perturbation theory (PT). In this work, we develop a PT-based theoretical model for \( P^{\mathrm{x}} \), building on existing PT models for \( P^{\mathrm{pre}}\) and \( P^{\mathrm{post}} \) \citep{hikage2017perturbation,Hikage:2019ihj}.

The structure of the paper is as follows. In Sec.~\ref{sec:model}, we develop the theoretical model for \( P^{\mathrm{x}} \), which constitutes the main result of this work. Sec.~\ref{sec:validation} presents the validation of the model using high-fidelity \( N \)-body simulations. Finally, we summarize our findings and conclusions in Sec.~\ref{sec:conclusion}.

\newpage
\section{A theory model for the cross-power spectrum}
\label{sec:model}

In this section, we present the formalism for the cross-power spectrum between pre- and post-reconstructed matter density fields in redshift space, developed up to the one-loop level within the framework of perturbation theory. This approach is motivated by previous studies on modeling the power spectrum of the post-reconstructed field \citep{hikage2017perturbation,Hikage:2019ihj}. Additionally, we discuss the asymptotic behavior of the one-loop corrections to provide further insight into their impact on the cross-power spectrum.

{When converting the observed redshift \(z\) to the comoving distance \(x\) in redshift space, two effects need to be considered: geometric distortions from the Alcock-Paczynski (AP) effect \citep{Alcock:1979mp}, which result from differences between the assumed fiducial cosmology and the true cosmology, and redshift-space distortions (RSD) \citep{Kaiser:1987qv} induced by the peculiar velocities of galaxies.
Neglecting the peculiar velocity of the observer, we further assume that the scales of interest are much smaller than the curvature scale of the Universe, allowing us to adopt a local Cartesian coordinate system in the survey volume. 
Under these assumptions, the initial Lagrangian position \(\mathbf{q}\) of a galaxy can be mapped to its Eulerian comoving position \(\mathbf{x}\) {at a time \(t\)} in redshift space through a displacement field \(\mathbf{\Psi}^{\mathrm{z}}(\mathbf{q}{,t})\), expressed as
\begin{equation}
    \mathbf{x}{(\mathbf{q},t)} = \mathbf{q} + \mathbf{\Psi}^{\mathrm{z}}(\mathbf{q}{,t}),
    \label{eq:xz}
\end{equation}
with the relation between the redshift-space displacement, \(\mathbf{\Psi}^{\mathrm{z}}\), and its real-space counterpart, \(\mathbf{\Psi}\), given by \citep[e.g.,][]{Matsubara_2008}
\begin{equation}
    \mathbf{\Psi}^{\mathrm{z}}(\mathbf{q},t) = \mathbf{\Psi}(\mathbf{q},t)+\frac{\mathbf{v}\cdot\hat{\mathbf{z}}}{aH}\hat{\mathbf{z}},
    \label{eq:psz}
\end{equation}
where \(\hat{\mathbf{z}}\) is the unit vector along the line of sight, \(a\) is the scale factor, \(H\) is the Hubble expansion rate, and \(\mathbf{v}=a\frac{\mathrm{d}\mathbf{\Psi}}{\mathrm{d}t} \equiv a\dot{\mathbf{\Psi}}\) represents the peculiar velocity. Throughout this paper, we assume the velocity field is irrotational and adopt the distant observer approximation, treating \(\hat{\mathbf{z}}\) as a fixed vector.
}

As a fundamental quantity in the Lagrangian description of the large-scale structure of the Universe, the displacement field {at redshift \(z\), \(\mathbf{\Psi}^{\mathrm{z}}(\mathbf{q},z)\), } can be evaluated using Lagrangian Perturbation Theory (LPT) {\citep[e.g.,][]{buchert1989class,hivon1994redshift,Bouchet:1994xp,catelan1995lagrangian,Catelan:1996hw,Bernardeau:2001qr,Matsubara_2008,White:2015eaa}}, yielding
\begin{equation}
\mathbf{\Psi}^{\mathrm{z}}(\mathbf{q}{,z}) = \sum_{n=1}^{\infty} \mathbf{\Psi}^{\mathrm{z}(n)}(\mathbf{q}{,z}),
\end{equation}
where \(\mathbf{\Psi}^{\mathrm{z}(n)}(\mathbf{q}{,z})\) represents the \(n\)-th order perturbative contribution.
In an Einstein–de Sitter Universe, {the temporal evolution of displacement perturbations can be factored out from the perturbation kernel, resulting in \(\mathbf{\Psi}^{(n)}(z) \propto D^n(z)\), where \(D(z)\) is the linear growth factor {normalized by \( D(z=0)=1 \)}. By further incorporating Eqs.~(\ref{eq:xz}) and ~(\ref{eq:psz}),
one can derive the transformation from the real-space displacement perturbation \(\mathbf{\Psi}^{(n)}\) to the redshift-space displacement perturbation \(\mathbf{\Psi}^{\mathrm{z}(n)}\) as
\begin{equation}
    \mathbf{\Psi}^{\mathrm{z}(n)} = \mathbf{R}^{(n)}\mathbf{\Psi}^{(n)},
\end{equation}
where \(\mathbf{R}^{(n)}\) is the redshift-space distortion tensor, defined as
\begin{equation}
R_{ij}^{(n)} = \delta_{ij} + n f \hat{z}_i \hat{z}_j,
\end{equation}
with the logarithmic linear growth rate \(f \equiv \mathrm{d} \ln D / \mathrm{d} \ln a\). And in this case,} the \(n\)-th order displacement perturbation in Fourier space is given by {\citep{Matsubara_2008}}
\begin{equation}
\mathbf{\tilde{\Psi}}_{\mathbf{k}}^{\mathrm{z}(n)} {\left(z\right)} = \frac{i D^{n}(z)}{n!} \int \frac{\mathrm{d}^3 \mathbf{k}_1 \cdots \mathrm{d}^3 \mathbf{k}_n}{(2\pi)^{3n-3}} \, \delta_{\mathrm{D}}\left(\sum_{j=1}^{n} \mathbf{k}_j - \mathbf{k}\right) 
\times \mathbf{L}^{\mathrm{z}(n)}(\mathbf{k}_1, \dots, \mathbf{k}_n) \, \tilde{\delta}^{\mathrm{L}}_{\mathbf{k}_1} \cdots \tilde{\delta}^{\mathrm{L}}_{\mathbf{k}_n}, 
\label{eq:psi}
\end{equation}
{where \(\delta_{\mathrm{D}}\) is the Dirac delta function, and \(\tilde{\delta}^{\mathrm{L}}_{\mathbf{k}}\) denotes the linear density contrast in Fourier space at redshift \(z = 0\).
The redshift-space LPT kernel \(\mathbf{L}^{\mathrm{z}(n)}\) is related to its real-space counterpart \(\mathbf{L}^{(n)}\) through the mapping
\begin{equation}
\mathbf{L}^{\mathrm{z}(n)} = \mathbf{R}^{(n)} \, \mathbf{L}^{(n)}.
\end{equation}}
The explicit forms of the LPT kernel \(\mathbf{L}^{(n)}\) for \(n = 1, 2, 3\) have been extensively studied in the literature {\citep[e.g.,][]{catelan1995lagrangian, Matsubara_2008, hikage2017perturbation}}.

Assuming that the initial distribution of galaxies is sufficiently uniform, the redshift-space mass overdensity, derived from the continuity equation, can be expressed as
\begin{equation}
\delta^{\mathrm{z}}(\mathbf{x}) = \int \mathrm{d}^3 \mathbf{q} \, \delta_{\mathrm{D}} \left[ \mathbf{x} - \mathbf{q} - \mathbf{\Psi}^{\mathrm{z}}(\mathbf{q}) \right] - 1,
\end{equation}
where \(\mathbf{\Psi}^{\mathrm{z}}(\mathbf{q})\) is the redshift-space displacement field.

Taking the Fourier transform of the overdensity field yields
\begin{equation}
\tilde{\delta}^{\mathrm{z}}(\mathbf{k}) = \int \mathrm{d}^3 \mathbf{q} \, e^{-i \mathbf{k} \cdot \mathbf{q}} \left( e^{-i \mathbf{k} \cdot \mathbf{\Psi}^{\mathrm{z}}(\mathbf{q})} - 1 \right),
\end{equation}
which describes the redshift-space overdensity in Fourier space as a function of the wavevector \(\mathbf{k}\).

This formulation establishes a connection between the unknown displacement field and the observed density field in redshift space. By utilizing the perturbative expansion of the displacement field in terms of the linear density field, as shown in Eq.~(\ref{eq:psi}), we can relate the perturbations of the density field to those of the displacement field. 

Within the framework of Standard Perturbation Theory (SPT), the redshift-space density field can be expanded as a series of perturbative terms. The \(n\)-th order perturbation is given by
\begin{equation}
\tilde{\delta}^{\mathrm{z}(n)}(\mathbf{k}) = D^{n}(z) \int \frac{\mathrm{d}^3 \mathbf{k}_1 \cdots \mathrm{d}^3 \mathbf{k}_n}{(2\pi)^{3n-3}} \, \delta_{\mathrm{D}}\left(\sum_{j=1}^{n} \mathbf{k}_j - \mathbf{k}\right) 
\times F_{n}^{\mathrm{z}}(\mathbf{k}_1, \dots, \mathbf{k}_n) \, \tilde{\delta}^{\mathrm{L}}_{\mathbf{k}_1} \cdots \tilde{\delta}^{\mathrm{L}}_{\mathbf{k}_n}, 
\label{eq:dkz}
\end{equation}
where \(F_{n}^{\mathrm{z}}\) is the Eulerian perturbation kernel in redshift space, \(D(z)\) is the linear growth factor, and \(\delta_{\mathrm{D}}\) is the Dirac delta function ensuring momentum conservation. For \(n = 1, 2, 3\), the kernels \(F_{n}^{\mathrm{z}}\) can be explicitly constructed using the Lagrangian Perturbation Theory (LPT) kernels \citep{Matsubara_2011, Hikage:2020fte}.

The initial conditions of the density field have evolved over cosmic time under complex nonlinear dynamics. In principle, this evolution can be partially reversed to recover a more linear representation of the present-day galaxy density field. Following the standard reconstruction algorithm \citep{Eisenstein:2006nk}, the bulk flow motions can be corrected by computing the shift field, defined as
\begin{equation}
\mathbf{s}^{\mathrm{z}}(\mathbf{x}) = \int \frac{\mathrm{d}^3 \mathbf{k}}{(2\pi)^3} \, \tilde{\mathbf{s}}^{\mathrm{z}}_{\mathbf{k}} \, e^{i \mathbf{k} \cdot \mathbf{x}},
\end{equation}
where the shift field in Fourier space, \(\tilde{\mathbf{s}}^{\mathrm{z}}_{\mathbf{k}}\), is calculated based on the Zel'dovich approximation \citep{zel1970gravitational}:
\begin{equation}
\tilde{\mathbf{s}}^{\mathrm{z}}_{\mathbf{k}} = -i \, W(k) \, \mathbf{L}^{(1)}(\mathbf{k}) \, \tilde{\delta}^{\mathrm{z}}_{\mathbf{k}}.
\label{eq:sz}
\end{equation}
Here, \(W(k) = \exp\left(-\frac{k^2 R_{\mathrm{s}}^2}{2}\right)\) is a Gaussian smoothing kernel with a characteristic smoothing scale \(R_{\mathrm{s}}\), designed to suppress high-\(k\) (small-scale) modes. This smoothing effectively isolates large-scale bulk flows, providing a linear estimate of the negative displacement field necessary for the reconstruction process.

{There are different conventions for handling RSD in reconstruction \citep[e.g.,][]{White:2015eaa,Cohn:2015ljb,Seo_2016,Chen_2019}. The method employed in this work can be viewed as a special case of either Rec-Sym or Rec-Iso, where the definitions of Rec-Sym and Rec-Iso can be found in the literature \citep[e.g.,][]{Chen_2019}. We directly estimate the shift field in redshift space as Eq.~(\ref{eq:sz}), and in practice, both data and random particles are applied the same shift field. Interpreting within the framework of Rec-Sym \citep{White:2015eaa}, this procedure indicates that we effectively assumed a linear growth rate of \(f_{\text{rec}} = 0\) during the reconstruction. Interestingly, under this assumption, Rec-Sym and Rec-Iso become equivalent.
Although the assumed \( f_{\text{rec}} \) deviates significantly from the true value, we have accurately accounted for this process in our modeling. Furthermore, this particular reconstruction approach offers a notable advantage by simplifying the computational complexity of the PT-based modeling. Unlike the case of Rec-Sym reconstruction with \( f_{\text{rec}} \neq 0 \), our approach avoids the additional line-of-sight dependence in the denominator of the shift field expression \citep{Zhao:2024xit}.}

By applying a perturbative expansion to \(\tilde{\delta}_{\mathbf{k}}^{\mathrm{z}}\), the \(n\)-th order perturbation of the shift field can be expressed as {\citep{Hikage:2019ihj}}
\begin{equation}
\tilde{\mathbf{s}}_{\mathbf{k}}^{\mathrm{z}(n)} = \frac{i D^{n}(z)}{n!} \int \frac{\mathrm{d}^3 \mathbf{k}_1 \cdots \mathrm{d}^3 \mathbf{k}_n}{(2\pi)^{3n-3}} \, \delta_{\mathrm{D}}\left(\sum_{j=1}^{n} \mathbf{k}_j - \mathbf{k}\right) 
\times \mathbf{S}^{\mathrm{z}(n)}(\mathbf{k}_1, \dots, \mathbf{k}_n) \, \tilde{\delta}^{\mathrm{L}}_{\mathbf{k}_1} \cdots \tilde{\delta}^{\mathrm{L}}_{\mathbf{k}_n},
\end{equation}
where \(\mathbf{S}^{\mathrm{z}(n)}(\mathbf{k}_1, \dots, \mathbf{k}_n)\) is the perturbation kernel of the shift field, defined as
\begin{equation}
\mathbf{S}^{\mathrm{z}(n)}(\mathbf{k}_1, \dots, \mathbf{k}_n) = -n! \, W(k) \, \mathbf{L}^{(1)}(\mathbf{k}) \, F_{n}^{\mathrm{z}}(\mathbf{k}_1, \dots, \mathbf{k}_n).
\end{equation}
Here, \(W(k)\) is the smoothing kernel, \(\mathbf{L}^{(1)}(\mathbf{k})\) is the first-order Lagrangian perturbation kernel, and \(F_{n}^{\mathrm{z}}\) represents the Eulerian perturbation kernel in redshift space. This formulation establishes a direct connection between the shift field and the underlying linear density perturbations through standard perturbation theory.

In the standard reconstruction algorithm, the shift field \(\mathbf{s}^{\mathrm{z}}(\mathbf{x})\) is applied to the original density field, which has evolved nonlinearly from the displacement field \(\mathbf{\Psi}^{\mathrm{z}}(\mathbf{q})\). This operation yields the displaced density field, expressed as
\begin{equation}
\tilde{\delta}_{\mathbf{k}}^{\mathrm{z}(\mathrm{d})} = \int \mathrm{d}^3 \mathbf{q} \, e^{-i \mathbf{k} \cdot \mathbf{q}} \left( e^{-i \mathbf{k} \cdot \left[\mathbf{\Psi}^{\mathrm{z}}(\mathbf{q}) + \mathbf{s}^{\mathrm{z}}(\mathbf{x})\right]} - 1 \right).
\end{equation}
The shift field partially cancels the linear, bulk-flow-dominated evolution of the density field, leaving the displaced density field primarily sensitive to the nonlinear components of the displacement field. On large scales, where the displacement field is dominated by its linear component, the relation \(\mathbf{\Psi}^{\mathrm{z}} + \mathbf{s}^{\mathrm{z}} \rightarrow \mathbf{0}\) holds, implying that particles are nearly returned to their original Lagrangian positions. In this regime, the displaced density field approximates the initial density field, enabling an effective reconstruction of the linear density modes.

To further isolate the linear information from the nonlinearly evolved density field, the shift field \(\mathbf{s}^{\mathrm{z}}(\mathbf{q})\) is also applied to a uniform density field in Lagrangian space, yielding the shifted density field:
\begin{equation}
\tilde{\delta}_{\mathbf{k}}^{\mathrm{z}(\mathrm{s})} = \int \mathrm{d}^3 \mathbf{q} \, e^{-i \mathbf{k} \cdot \mathbf{q}} \left( e^{-i \mathbf{k} \cdot \mathbf{s}^{\mathrm{z}}(\mathbf{q})} - 1 \right).
\end{equation}
This shifted density field contains information about the original nonlinear density fluctuations associated with linear displacements, as modeled by the Zel'dovich approximation. Since the uniformly distributed particles are displaced in directions opposite to those of the actual galaxies, the reconstructed density field is defined as the difference between the displaced and shifted density fields:
\begin{align}
\tilde{\delta}_{\mathbf{k}}^{\mathrm{z}(\mathrm{rec})} &\equiv \tilde{\delta}_{\mathbf{k}}^{\mathrm{z}(\mathrm{d})} - \tilde{\delta}_{\mathbf{k}}^{\mathrm{z}(\mathrm{s})} \nonumber \\
&= \int \mathrm{d}^3 \mathbf{q} \, e^{-i \mathbf{k} \cdot \mathbf{q}} \, e^{-i \mathbf{k} \cdot \mathbf{s}^{\mathrm{z}}(\mathbf{q})} \left( e^{-i \mathbf{k} \cdot \left[\mathbf{\Psi}^{\mathrm{z}}(\mathbf{q}) + \mathbf{s}^{\mathrm{z}}(\mathbf{x}) - \mathbf{s}^{\mathrm{z}}(\mathbf{q})\right]} - 1 \right).
\end{align}

The \(n\)-th order perturbation of the reconstructed density field can be written as
\begin{equation}
\tilde{\delta}_{\mathbf{k}}^{\mathrm{z}(\mathrm{rec}, n)} = D^{n}(z) \int \frac{\mathrm{d}^3 \mathbf{k}_1 \cdots \mathrm{d}^3 \mathbf{k}_n}{(2\pi)^{3n-3}} \, \delta_{\mathrm{D}}\left(\sum_{j=1}^{n} \mathbf{k}_j - \mathbf{k}\right) 
\times F_{n}^{\mathrm{z}(\mathrm{rec})}(\mathbf{k}_1, \dots, \mathbf{k}_n) \, \tilde{\delta}^{\mathrm{L}}_{\mathbf{k}_1} \cdots \tilde{\delta}^{\mathrm{L}}_{\mathbf{k}_n}, 
\label{eq:dkzr}
\end{equation}
where \(F_{n}^{\mathrm{z}(\mathrm{rec})}\) represents the post-reconstruction perturbation kernels, as derived in \citet{Hikage:2019ihj, Hikage:2020fte}.

The standard reconstruction algorithm decomposes the density field into two components: the linear Zel'dovich displacement and the nonlinear displacement originating from the evolved nonlinear density field. Different operations are applied to these components to achieve partial decoupling of the perturbation modes. The degree of coupling between these modes reflects the imprint of nonlinear information present in the density field before and after reconstruction. 

To analyze the cross-correlation between the pre- and post-reconstructed matter density fields, we define the cross-power spectrum in redshift space as
\begin{equation}
\left\langle \tilde{\delta}_{\mathbf{k}}^{\mathrm{z}} \, \tilde{\delta}_{\mathbf{k}^{\prime}}^{\mathrm{z}(\mathrm{rec})} \right\rangle \equiv (2\pi)^3 \delta_{\mathrm{D}}(\mathbf{k} + \mathbf{k}^{\prime}) P^{\mathrm{z}(\mathrm{x})}(\mathbf{k}),
\label{eq:pzxk}
\end{equation}
where \(P^{\mathrm{z}(\mathrm{x})}(\mathbf{k})\) is the redshift-space cross-power spectrum between the pre- and post-reconstructed density fields. 

Using Eqs.~(\ref{eq:dkz}), (\ref{eq:dkzr}), and (\ref{eq:pzxk}), we derive the redshift-space cross-power spectrum up to one-loop order (see Appendix~\ref{de_px} for details):
\begin{equation}
P_{\mathrm{1-loop}}^{\mathrm{z}(\mathrm{x})}(k, \mu) = D^2(z) P_{11}^{\mathrm{z}(\mathrm{x})}(k, \mu) + D^4(z) \left[ P_{22}^{\mathrm{z}(\mathrm{x})}(k, \mu) + P_{13}^{\mathrm{z}(\mathrm{x})}(k, \mu) \right],
\end{equation}
where the terms correspond to the following contributions:
\begin{align}
P_{11}^{\mathrm{z}(\mathrm{x})}(k, \mu) &= \left(1 + f \mu^2\right)^2 P_{\mathrm{L}}(k), \\
P_{22}^{\mathrm{z}(\mathrm{x})}(k, \mu) &= 2 \int \frac{\mathrm{d}^3 \mathbf{p}}{(2\pi)^3} \left[ F_{2}^{\mathrm{z}(\mathrm{x})}(\mathbf{k} - \mathbf{p}, \mathbf{p}) \right]^2 P_{\mathrm{L}}(|\mathbf{k} - \mathbf{p}|) P_{\mathrm{L}}(p), \label{eq:p22} \\
P_{13}^{\mathrm{z}(\mathrm{x})}(k, \mu) &= 6 \left(1 + f \mu^2\right) P_{\mathrm{L}}(k) \int \frac{\mathrm{d}^3 \mathbf{p}}{(2\pi)^3} F_{3}^{\mathrm{z}(\mathrm{x})}(\mathbf{k}, \mathbf{p}, -\mathbf{p}) P_{\mathrm{L}}(p), \label{eq:p13}
\end{align}
with \(P_{\mathrm{L}}(k)\) denoting the linear matter power spectrum, and \(f \equiv \mathrm{d} \ln D / \mathrm{d} \ln a\) representing the linear growth rate.

The effective perturbation kernels for the one-loop terms are defined as
\begin{align}
F_{2}^{\mathrm{z}(\mathrm{x})}(\mathbf{k} - \mathbf{p}, \mathbf{p}) &\equiv \sqrt{ F_{2}^{\mathrm{z}}(\mathbf{k} - \mathbf{p}, \mathbf{p}) \, F_{2}^{\mathrm{z}(\mathrm{rec})}(\mathbf{k} - \mathbf{p}, \mathbf{p}) }, \label{eq:f2} \\
F_{3}^{\mathrm{z}(\mathrm{x})}(\mathbf{k}, \mathbf{p}, -\mathbf{p}) &\equiv \frac{1}{2} \left[ F_{3}^{\mathrm{z}}(\mathbf{k}, \mathbf{p}, -\mathbf{p}) + F_{3}^{\mathrm{z}(\mathrm{rec})}(\mathbf{k}, \mathbf{p}, -\mathbf{p}) \right]. \label{eq:f3}
\end{align}

Since the reconstruction process does not modify the linear-order density perturbations, we have 
\[
P_{11}^{\mathrm{z}(\mathrm{x})} = P_{11}^{\mathrm{z}(\mathrm{rec})} = P_{11}^{\mathrm{z}(\mathrm{pre})}.
\]
Thus, the differences between the pre- and post-reconstructed power spectra arise solely from the nonlinear corrections, encapsulated in the \(P_{22}\) and \(P_{13}\) terms.

\begin{figure*}
\begin{center}
\includegraphics[width=9cm]{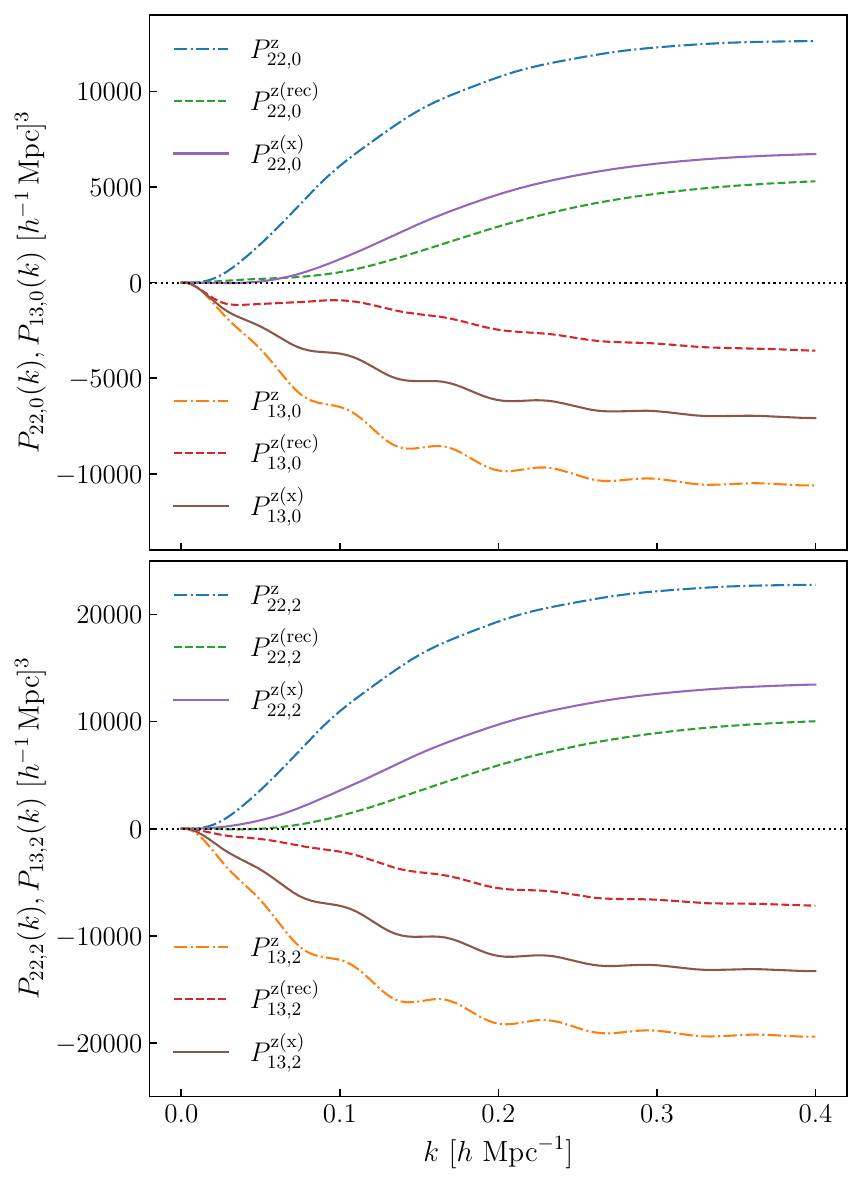}\includegraphics[width=9cm]{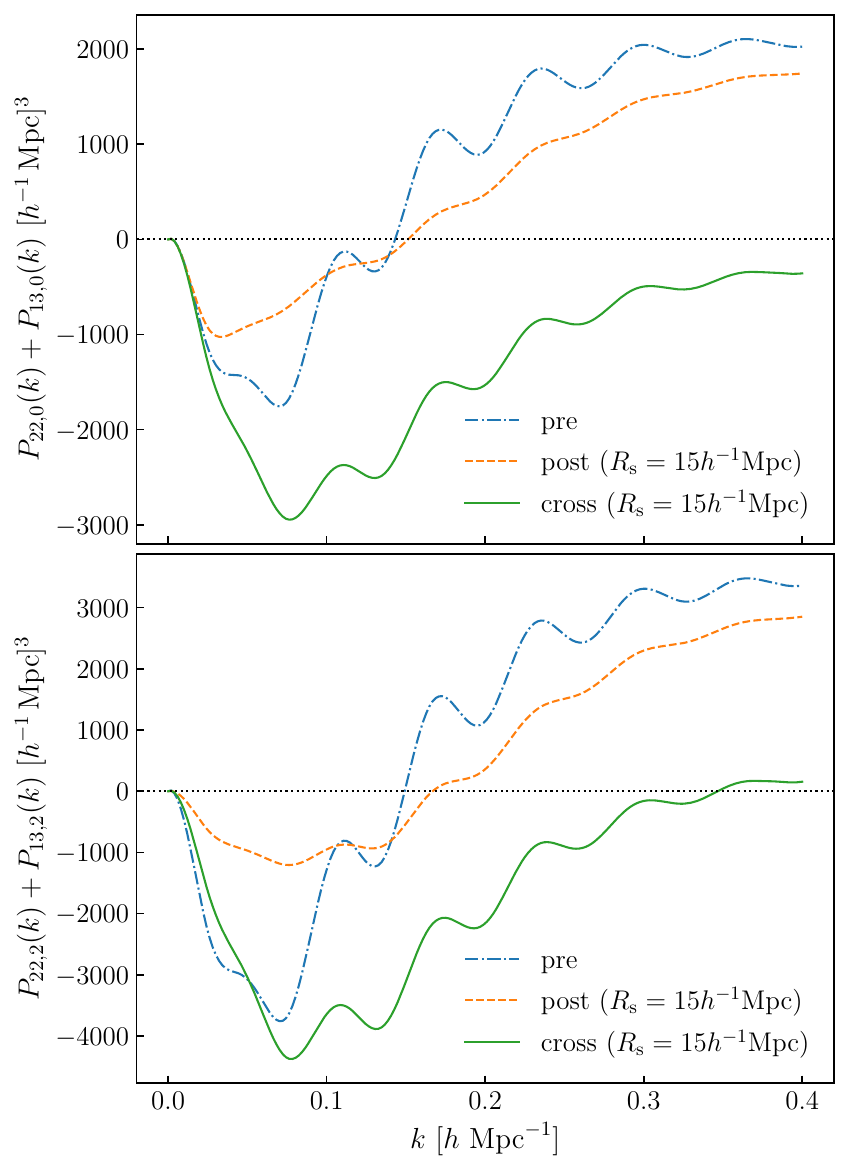}
\caption{\label{fig_1loop}Comparison of the monopole and quadrupole moments of the one-loop components for the pre-reconstruction, post-reconstruction, and cross matter power spectra, shown as dot-dashed, dashed, and solid lines, respectively. These theoretical predictions are computed based on Eq.~(\ref{eq:plk}), with \(P(k, \mu)\) replaced by the corresponding one-loop terms {\( P_{22}(k,\mu) \) and \( P_{13}(k,\mu) \)} from Standard Perturbation Theory (SPT). {And the linear power spectrum used to calculate the one-loop terms is given at redshift \( z = 0 \).} The growth rate parameter \(f\) and the linear power spectrum are consistent with the simulations described in Sec.~\ref{sec:validation}. A smoothing scale of \(R_{\mathrm{s}} = 15\, h^{-1}\mathrm{Mpc}\) is adopted for the reconstruction as an example.}
\end{center}
\end{figure*}

Since reconstruction partially transfers higher-order statistical information from the pre-reconstruction density field into the two-point statistics of the reconstructed density field, the post-reconstruction power spectrum can be expressed in a form that encapsulates contributions from higher-order statistics \citep{Schmittfull_2015, Wang:2022nlx, Sugiyama:2024ggt}. Eqs.~(\ref{eq:p22})–(\ref{eq:f3}) illustrate that the one-loop terms of the cross-power spectrum are constructed from both the pre- and post-reconstruction density perturbation kernels. As a result, this cross-power spectrum, as a two-point statistic, can potentially capture complementary higher-order statistical information beyond what is contained in the individual pre- and post-reconstruction power spectra.

{To compare our theoretical model with the measured data from simulations, we compute the matter power spectrum multipoles using the formulae for \(P(k, \mu)\) and the linear matter power spectrum generated by {\tt CAMB} \citep{Lewis:1999bs}. The power spectrum multipoles are calculated as:
\begin{equation}
P_{\ell}(k) = \frac{2\ell + 1}{2} \int_{-1}^{1} \mathrm{d}\mu \, P(k, \mu) \, \mathcal{L}_{\ell}(\mu), \label{eq:plk}
\end{equation} where \(\mathcal{L}_{\ell}(\mu)\) is the \(\ell\)-th order Legendre polynomial.}

Computing as Eq.~(\ref{eq:plk}), Figure~\ref{fig_1loop} compares the monopole and quadrupole moments of the one-loop terms for the pre-reconstruction, post-reconstruction, and cross-power spectra. {Note that the theoretical results for all three types of power spectra in Figure~\ref{fig_1loop} are computed using our implementation of the theoretical models. The theoretical models for the pre- and post-reconstruction power spectra are consistent with those in \citet{Hikage:2019ihj}, while the Fig. 1 of which adopts a different smoothing scale of \(R_{\mathrm{s}} = 10\, h^{-1}\mathrm{Mpc}\).} In the left panel, it is evident that the amplitudes of both one-loop components, \(P_{22}^{\mathrm{z}}\) and \(P_{13}^{\mathrm{z}}\), are suppressed after reconstruction. The monopole and quadrupole of the cross-power spectrum, \(P_{22}^{\mathrm{z}(\mathrm{x})}\) and \(P_{13}^{\mathrm{z}(\mathrm{x})}\), generally lie between the pre- and post-reconstruction cases. However, \(P_{22,0}^{\mathrm{z}(\mathrm{x})}\) exhibits a slightly smaller amplitude than \(P_{22,0}^{\mathrm{z}(\mathrm{rec})}\) on large scales. The right panel of Figure~\ref{fig_1loop} shows the total one-loop corrections for the different power spectra. Compared to the pre-reconstruction case, the BAO features in the one-loop terms of the post-reconstruction power spectrum are significantly smoothed, although they are partially preserved in the cross-power spectrum \(P^{\mathrm{z}(\mathrm{x})}\). Additionally, \(P_{22,2}^{\mathrm{z}(\mathrm{x})} + P_{13,2}^{\mathrm{z}(\mathrm{x})}\) is slightly larger than its pre-reconstruction counterpart on large scales. In contrast, the combined monopole and quadrupole components, \(P_{22,0}^{\mathrm{z}(\mathrm{x})} + P_{13,0}^{\mathrm{z}(\mathrm{x})}\) and \(P_{22,2}^{\mathrm{z}(\mathrm{x})} + P_{13,2}^{\mathrm{z}(\mathrm{x})}\), are generally smaller than those in both the pre- and post-reconstruction cases. Notably, while \(P_{22}^{\mathrm{z}} \sim -P_{13}^{\mathrm{z}}\) and \(P_{22}^{\mathrm{z}(\mathrm{rec})} \sim -P_{13}^{\mathrm{z}(\mathrm{rec})}\), we observe \(P_{22}^{\mathrm{z}(\mathrm{x})} \lesssim -P_{13}^{\mathrm{z}(\mathrm{x})}\). As a result, the monopole and quadrupole of the cross-power spectrum exhibit a more rapid decay with increasing wavenumber compared to the pre- and post-reconstruction cases. This behavior is primarily attributed to the absence of infrared (IR) cancellation in the cross-power spectrum \citep{Sugiyama:2024eye}, indicating a degree of decorrelation between the matter density fields before and after reconstruction, consistent with findings from simulations \citep{Wang:2022nlx}.

\begin{figure*}
\begin{center}
\includegraphics[width=18cm]{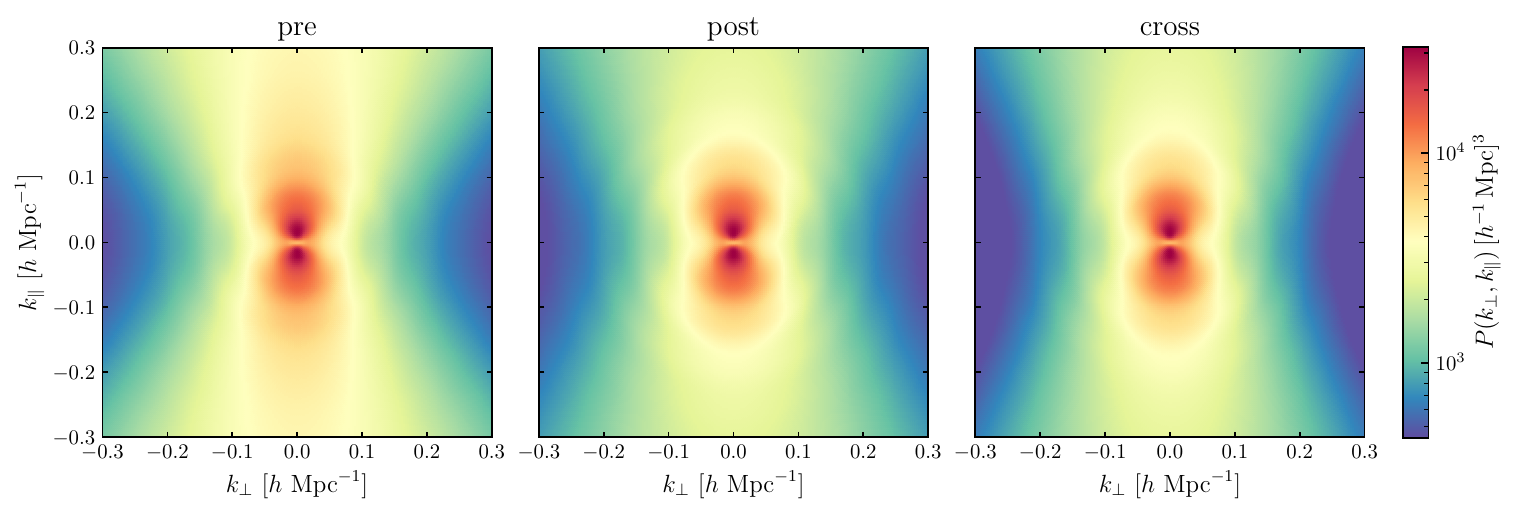}
\caption{\label{fig_pkmu}Comparison of three types of two-dimensional matter power spectra in redshift space, computed using the Standard Perturbation Theory (SPT) model up to one-loop order. The calculations are performed with the same cosmological parameters as those used in the simulations for model validation in Sec.~\ref{sec:validation}. The reconstruction adopts a smoothing scale of \(R_{\mathrm{s}} = 15\, h^{-1}\mathrm{Mpc}\), and the results correspond to a redshift of \(z = 1.02\).}
\end{center}
\end{figure*}

To better illustrate the differences between the cross-power spectrum and the pre- and post-reconstruction power spectra, we present the two-dimensional power spectra predicted by the one-loop Standard Perturbation Theory (SPT) in Figure~\ref{fig_pkmu}. The power spectra are noticeably enhanced along the line of sight due to redshift-space distortions (RSD), with the effect being particularly pronounced in the pre-reconstruction case at small scales (i.e., large \(k_{\parallel}\)). Following reconstruction, the Baryon Acoustic Oscillation (BAO) signal becomes more prominent, as expected. The cross-power spectrum exhibits a stronger BAO feature compared to the pre-reconstruction case but remains slightly weaker than that of the post-reconstruction spectrum. At small scales perpendicular to the line of sight (i.e., large \(k_{\perp}\)), the cross-power spectrum shows a significant suppression relative to both the pre- and post-reconstruction power spectra.

To better understand the properties of the one-loop power spectrum, it is instructive to analyze its asymptotic behavior in the limits of large and small external momenta. We first consider the high-\(k\) limit, i.e., \(k \gg p\), to investigate the infrared (IR) effects on \(P^{\mathrm{z}(\mathrm{x})}\). In this regime, the second- and third-order perturbation kernels for the pre- and post-reconstructed matter density fields can be approximated as \citep{Sugiyama:2024eye}:
\begin{equation}
F_{2,\mathrm{IR}}^{\mathrm{z}}(\mathbf{k}, \mathbf{p}) = \frac{1}{2} \, \mathbf{k} \cdot \mathbf{L}^{\mathrm{z}(1)}(\mathbf{p}) \, F_{1}^{\mathrm{z}}(\mathbf{k}),
\end{equation}
\begin{equation}
F_{3,\mathrm{IR}}^{\mathrm{z}}(\mathbf{k}, \mathbf{p}, -\mathbf{p}) = -\frac{1}{3!} \left[\mathbf{k} \cdot \mathbf{L}^{\mathrm{z}(1)}(\mathbf{p})\right]^2 F_{1}^{\mathrm{z}}(\mathbf{k}),
\end{equation}
\begin{equation}
F_{2,\mathrm{IR}}^{\mathrm{z}(\mathrm{rec})}(\mathbf{k}, \mathbf{p}) = \frac{1}{2} \left[\mathbf{k} \cdot \mathbf{L}^{\mathrm{z}(1)}(\mathbf{p}) + \mathbf{k} \cdot \mathbf{S}^{\mathrm{z}(1)}(\mathbf{p})\right] F_{1}^{\mathrm{z}}(\mathbf{k}),
\end{equation}
\begin{equation}
F_{3,\mathrm{IR}}^{\mathrm{z}(\mathrm{rec})}(\mathbf{k}, \mathbf{p}, -\mathbf{p}) = -\frac{1}{3!} \left[\mathbf{k} \cdot \mathbf{L}^{\mathrm{z}(1)}(\mathbf{p}) + \mathbf{k} \cdot \mathbf{S}^{\mathrm{z}(1)}(\mathbf{p})\right]^2 F_{1}^{\mathrm{z}}(\mathbf{k}).
\end{equation}

Using these approximations, the high-\(k\) limit of the cross-power spectrum is given by:
\begin{align}
P_{22}^{\mathrm{z}(\mathrm{x})}(\mathbf{k}) &\xrightarrow{k \gg p,\, |\mathbf{k} - \mathbf{p}|} 
2 \times 2 \int \frac{\mathrm{d}^3 \mathbf{p}}{(2\pi)^3} \, F_{2,\mathrm{IR}}^{\mathrm{z}}(\mathbf{k}, \mathbf{p}) F_{2,\mathrm{IR}}^{\mathrm{z}(\mathrm{rec})}(\mathbf{k}, \mathbf{p}) P_{\mathrm{L}}(p) P_{\mathrm{L}}(k) \nonumber \\
&= P_{\mathbf{\Psi}^2}(\mathbf{k}) + P_{\mathbf{\Psi}, \mathbf{s}}(\mathbf{k}),
\end{align}
\begin{align}
P_{13}^{\mathrm{z}(\mathrm{x})}(\mathbf{k}) &\xrightarrow{k \gg p} 
6 F_{1}^{\mathrm{z}}(\mathbf{k}) P_{\mathrm{L}}(k) \int \frac{\mathrm{d}^3 \mathbf{p}}{(2\pi)^3} \, \frac{1}{2} \left[F_{3,\mathrm{IR}}^{\mathrm{z}}(\mathbf{k}, \mathbf{p}, -\mathbf{p}) + F_{3,\mathrm{IR}}^{\mathrm{z}(\mathrm{rec})}(\mathbf{k}, \mathbf{p}, -\mathbf{p})\right] P_{\mathrm{L}}(p) \nonumber \\
&= -P_{\mathbf{\Psi}^2}(\mathbf{k}) - P_{\mathbf{\Psi}, \mathbf{s}}(\mathbf{k}) - \frac{1}{2} P_{\mathbf{s}^2}(\mathbf{k}),
\end{align}
where the following definitions are used:
\begin{equation}
P_{\mathbf{\Psi}^2}(\mathbf{k}) \equiv \left[F_{1}^{\mathrm{z}}(\mathbf{k})\right]^2 P_{\mathrm{L}}(k) \int \frac{\mathrm{d}^3 \mathbf{p}}{(2\pi)^3} \, \left[\mathbf{k} \cdot \mathbf{L}^{\mathrm{z}(1)}(\mathbf{p})\right]^2 P_{\mathrm{L}}(p),
\end{equation}
\begin{equation}
P_{\mathbf{\Psi}, \mathbf{s}}(\mathbf{k}) \equiv \left[F_{1}^{\mathrm{z}}(\mathbf{k})\right]^2 P_{\mathrm{L}}(k) \int \frac{\mathrm{d}^3 \mathbf{p}}{(2\pi)^3} \, \left[\mathbf{k} \cdot \mathbf{L}^{\mathrm{z}(1)}(\mathbf{p}) \, \mathbf{k} \cdot \mathbf{S}^{\mathrm{z}(1)}(\mathbf{p})\right] P_{\mathrm{L}}(p),
\end{equation}
\begin{equation}
P_{\mathbf{s}^2}(\mathbf{k}) \equiv \left[F_{1}^{\mathrm{z}}(\mathbf{k})\right]^2 P_{\mathrm{L}}(k) \int \frac{\mathrm{d}^3 \mathbf{p}}{(2\pi)^3} \, \left[\mathbf{k} \cdot \mathbf{S}^{\mathrm{z}(1)}(\mathbf{p})\right]^2 P_{\mathrm{L}}(p).
\end{equation}

We observe that the contributions from the displacement field \(\mathbf{\Psi}^{\mathrm{z}}\) and the coupling term between \(\mathbf{\Psi}^{\mathrm{z}}\) and the shift field \(\mathbf{s}^{\mathrm{z}}\) cancel out when summing \(P_{22}^{\mathrm{z}(\mathrm{x})}\) and \(P_{13}^{\mathrm{z}(\mathrm{x})}\) in the IR limit. However, the term \(P_{\mathbf{s}^2}(\mathbf{k})\) breaks this IR cancellation in the one-loop cross-matter power spectrum, leaving residual information from the shift field in the high-\(k\) limit of \(P_{\mathrm{1-loop}}^{\mathrm{z}(\mathrm{x})}\). This absence of IR cancellation results in an exponential decay of the cross-matter power spectrum at high \(k\), as highlighted in \citep{Sugiyama:2024eye}, based on a non-perturbative treatment of IR effects.

In the low-\(k\) limit, we set \(f = 0\) to explore how the one-loop terms of the cross-power spectrum are influenced by ultraviolet (UV) modes in real space. Specifically, we focus on the leading-order contributions from the reconstruction, defined as \(\Delta P_{22}^{(\mathrm{x})} \equiv P_{22}^{(\mathrm{x})} - P_{22}\) and \(\Delta P_{13}^{(\mathrm{x})} \equiv P_{13}^{(\mathrm{x})} - P_{13}\).

After performing an asymptotic expansion of the one-loop integrals in the \(k \ll p\) limit and neglecting higher-order terms, we obtain:
\begin{equation}
P_{22}^{(\mathrm{x})}(k) \xrightarrow{k \ll p} P_{22, \mathrm{UV}}(k) + \Delta P_{22, \mathrm{UV}}^{(\mathrm{x})}(k),
\end{equation}
\begin{equation}
P_{13}^{(\mathrm{x})}(k) \xrightarrow{k \ll p} P_{13, \mathrm{UV}}(k) + \Delta P_{13, \mathrm{UV}}^{(\mathrm{x})}(k),
\end{equation}
where
\begin{equation}
P_{22, \mathrm{UV}}(k) = \frac{9}{196 \pi^2} k^4 \int_{k \ll p} \mathrm{d}p \, p^2 \frac{P_{\mathrm{L}}^2(p)}{p^4},
\end{equation}
\begin{equation}
P_{13, \mathrm{UV}}(k) = -k^2 P_{\mathrm{L}}(k) \int_{k \ll p} \mathrm{d}p \, p^2 \frac{P_{\mathrm{L}}(p)}{p^2} \left( \frac{61}{630 \pi^2} - \frac{2 k^2}{105 \pi^2 p^2} \right),
\end{equation}
represent the UV limits of the pre-reconstruction terms \(P_{22}\) and \(P_{13}\), which are the dominant components of \(P_{\mathrm{UV}}^{(\mathrm{x})}\) \citep{Baldauf:2014qfa}.

The reconstruction contributions are given by:
\begin{equation}
\Delta P_{22, \mathrm{UV}}^{(\mathrm{x})}(k) = \frac{1}{4\pi^2} k^4 \int_{k \ll p} \frac{\mathrm{d}p}{p^2} P_{\mathrm{L}}^2(p) \left[ -\frac{1}{21} - \frac{\cosh(k p R_{\mathrm{s}}^2)}{k^2 R_{\mathrm{s}}^2} e^{-\frac{1}{2} k^2 R_{\mathrm{s}}^2} \right] e^{-\frac{1}{2} p^2 R_{\mathrm{s}}^2}, \label{eq:p22uv}
\end{equation}
\begin{equation}
\Delta P_{13, \mathrm{UV}}^{(\mathrm{x})}(k) = -k^2 P_{\mathrm{L}}(k) \int_{k \ll p} \mathrm{d}p \, p^2 \frac{P_{\mathrm{L}}(p)}{p^2} \frac{p}{4 \pi^2 k^3 R_{\mathrm{s}}^2} e^{-\frac{1}{2} (k - p)^2 R_{\mathrm{s}}^2}, \label{eq:p13uv}
\end{equation}
which correspond to the UV limits of \(\Delta P_{22}^{(\mathrm{x})}\) and \(\Delta P_{13}^{(\mathrm{x})}\), respectively. These terms represent the leading-order UV contributions from the reconstruction.

The dominant UV contributions from \(\Delta P_{22}^{(\mathrm{x})}\) and \(\Delta P_{13}^{(\mathrm{x})}\) exhibit different \(k\)-dependencies compared to the pre-reconstruction case. In the limit \(R_{\mathrm{s}} \to +\infty\), the shift field vanishes, causing \(\Delta P_{22, \mathrm{UV}}^{(\mathrm{x})}\) and \(\Delta P_{13, \mathrm{UV}}^{(\mathrm{x})}\) to approach zero, as the system reverts to the pre-reconstruction state. Conversely, as \(R_{\mathrm{s}} \to 0\) (corresponding to an extreme reconstruction limit), these terms become divergent due to the strong influence of small-scale modes. Assuming the integrals in Eqs.~(\ref{eq:p22uv}) and (\ref{eq:p13uv}) converge, we find the following scaling behavior in the large-scale limit (\(k \to 0\)):
\[
\Delta P_{22, \mathrm{UV}}^{(\mathrm{x})} \propto k^2, \quad \Delta P_{13, \mathrm{UV}}^{(\mathrm{x})} \propto -k^{-1} P_{\mathrm{L}}(k).
\]
This implies that \(\Delta P_{13, \mathrm{UV}}^{(\mathrm{x})}\) may yield a non-vanishing contribution if \(P_{\mathrm{L}}(k) \propto k^n\) with \(n \leq 1\) as \(k \to 0\). However, it is important to note that Eqs.~(\ref{eq:p22uv}) and (\ref{eq:p13uv}) describe next-to-leading-order corrections compared to the dominant pre-reconstruction contributions. The UV effects of the reconstruction terms are further suppressed by the exponential damping factors in the integrands of Eqs.~(\ref{eq:p22uv}) and (\ref{eq:p13uv}). Therefore, focusing solely on the UV physics of the pre-reconstruction components and incorporating results from the Effective Field Theory of Large-Scale Structure (EFTofLSS) provides a reliable approximation for modeling the cross-power spectrum at the one-loop level.

\section{Validation Using High-Resolution Simulations}
\label{sec:validation}

To validate the performance of our theoretical model in constraining cosmological parameters, such as the linear growth rate, we employ high-resolution dark matter \(N\)-body simulations {which are generated using the publicly available TreePM code {\tt GADGET-2} \citep{Springel:2005mi}. These simulations were previously used in \citep{Hikage:2020fte,Wang:2022nlx}.}

The dataset consists of $4,000$ realizations of the same cosmological model at redshift \(z = 1.02\), based on the best-fit cosmological parameters from the Planck 2015 TT, TE, EE+lowP measurements \citep{Planck:2015fie}: \(\Omega_{b} = 0.0492\), \(\Omega_{m} = 0.3156\), \(h = 0.6727\), \(n_{s} = 0.9645\), and \(\sigma_{8} = 0.831\). Each simulation is performed in a cubic box with a side length of \(L = 500\, h^{-1} \mathrm{Mpc}\), containing \(512^3\) dark matter particles. {The density field are computed by assigning these particles to \(512^3\) grid cells using the clouds-in-cell (CIC) method. We use a fast Fourier transform (FFT)-based code \footnote{\url{https://github.com/chiaki-hikage/reconstruct_densityfield}} \citep{Hikage:2019ihj,Hikage:2020fte} to perform density field reconstruction and measure the power spectra after correcting the pixel window effect \citep[e.g.,][]{Jing:2004fq}. Additionally, periodic boundary conditions (PBC) are applied. For more details on the reconstruction and power spectrum measurements, please refer to Chiaki Hikage’s publicly available code and \citet{Hikage:2020fte}.} 

The standard reconstruction algorithm is applied to each mock catalog in redshift space, using smoothing scales of \(R_{\mathrm{s}} = 10, 15,\) and \(20\, h^{-1} \mathrm{Mpc}\). {In this work, we did not adopt the iterative reconstruction approach \citep[e.g.,][]{Seo:2009fp,Hada:2018fde,Schmittfull:2017uhh,Seo:2021nev,Ota:2021caz,Chen:2023iia}.} From these catalogs, we measure the monopole, quadrupole, and hexadecapole moments of the cross-power spectrum between the pre- and post-reconstructed matter density fields. However, due to significant uncertainties in both the modeling and measurement of the hexadecapole, we restrict our analysis to the monopole and quadrupole moments.

{Due to the small box size of the simulation thus a deficit large-scale modes}, the measured multipoles exhibit a sawtooth pattern on large scales, particularly in the quadrupole. To correct for these large-scale fluctuations, we incorporate power spectrum measurements from eight additional realizations with a larger simulation box of \(L = 4\, h^{-1} \mathrm{Gpc}\), generated using the same cosmological parameters. {In each realization, \(4096^3\) mass particles are assigned to \(2048^3\) grid cells using the CIC method \citep{Hikage:2020fte}.} The corrected power spectrum is given by:
\begin{equation}
P_{\ell}^{(\mathrm{x}),\,4\,h^{-1}\mathrm{Gpc}} = \frac{P_{\ell}^{(\mathrm{pre}),\,4\,h^{-1}\mathrm{Gpc}}}{P_{\ell}^{(\mathrm{pre}),\,500\,h^{-1}\mathrm{Mpc}}} \, P_{\ell}^{(\mathrm{x}),\,500\,h^{-1}\mathrm{Mpc}}, 
\label{eq:grid-corr}
\end{equation} where \(P_{\ell}^{(\mathrm{x}),\,500\,h^{-1}\mathrm{Mpc}}\) is the power spectrum multipole from the smaller box, and \(P_{\ell}^{(\mathrm{pre}),\,4\,h^{-1}\mathrm{Gpc}}\) and \(P_{\ell}^{(\mathrm{pre}),\,500\,h^{-1}\mathrm{Mpc}}\) are the pre-reconstruction power spectra from the larger and smaller simulation boxes, respectively. This correction ensures that large-scale modes are accurately represented in the final analysis.

To estimate the covariance matrix, we use the measurements from $4,000$ realizations of the dark matter \(N\)-body simulations. The covariance matrix is computed as:
\begin{equation}
\mathbf{Cov}_{\ell \ell^{\prime}}(k, k^{\prime}) = \frac{1}{N - 1} \sum_{i=1}^{N} \left[ P_{\ell, i}(k) - \bar{P}_{\ell}(k) \right] \left[ P_{\ell^{\prime}, i}(k^{\prime}) - \bar{P}_{\ell^{\prime}}(k^{\prime}) \right], \label{eq:cov}
\end{equation} where \(N = 4000\) is the number of realizations, and \(\bar{P}_{\ell}(k) = (1/N) \sum_{i=1}^{N} P_{\ell, i}(k)\) is the average power spectrum across all realizations. To obtain an unbiased estimator for the inverse covariance matrix, we apply the Hartlap correction factor \citep{Hartlap:2006kj}:
\begin{equation}
\mathbf{Cov}^{-1}_{\mathrm{corrected}} = \eta \, \mathbf{Cov}^{-1}, \quad \text{with} \quad \eta = \frac{N - N_{\mathrm{bin}} - 2}{N - 1},
\end{equation} where \(N_{\mathrm{bin}}\) is the number of \(k\)-bins used in the analysis. This correction accounts for the finite number of realizations and ensures accurate parameter estimation.

To ensure the practical applicability of our one-loop Standard Perturbation Theory (SPT) model, we incorporate counterterms from the Effective Field Theory (EFT) of Large-Scale Structure. These counterterms parameterize the impact of small-scale, non-perturbative physics that cannot be accurately captured by SPT alone. As discussed earlier, the dominant ultraviolet (UV) contributions at the one-loop level primarily arise from the pre-reconstruction components of \(P^{(\mathrm{x})}\). Therefore, adopting the same form of counterterms as used in the pre-reconstruction case is a reasonable approximation.

When fitting the linear growth rate \(f\) to the simulation data, we include the lowest-order EFT counterterms following the approach of \citet{Hikage:2019ihj, Hikage:2020fte}:
\begin{equation}
P_{\ell}^{\mathrm{theory}}(k) = P_{\ell}^{\mathrm{1-loop}}(k) + \alpha_{\ell} k^{2} P_{\ell}^{\mathrm{L}}(k), \label{eq:ct1}
\end{equation} where \(\alpha_{\ell}\) are free parameters that absorb the UV contributions. This form can be related to the counterterm expression in \citet{Perko:2016puo}:
\begin{equation}
P_{\ell}^{\mathrm{ct}}(k, \mu) = 2 \left( c_{\mathrm{ct}} + c_{r,1} \mu^{2} + c_{r,2} \mu^{4} \right) \left(1 + f \mu^{2}\right) k^{2} P_{\mathrm{L}}(k),
\end{equation} where \(c_{\mathrm{ct}}\), \(c_{r,1}\), and \(c_{r,2}\) are EFT coefficients that capture the small-scale physics. The relationships between \(\alpha_{\ell}\) and these coefficients are given by:
\begin{align}
\alpha_{0} &= \frac{14(3f + 5) c_{r,1} + 6(5f + 7) c_{r,2} + 70(f + 3) c_{\mathrm{ct}}}{7(3f^{2} + 10f + 15)}, \\
\alpha_{2} &= \frac{(6f + 7) c_{r,1} + (5f + 6) c_{r,2} + 7f c_{\mathrm{ct}}}{f(3f + 7)}, \\
\alpha_{4} &= \frac{22f c_{r,1} + 2(15f + 11) c_{r,2}}{11 f^{2}}.
\end{align}

It is important to note that our analysis does not explicitly include IR resummation. \citet{Sugiyama:2024eye} showed that for the cross-power spectrum, due to the lack of IR cancellation, a damping term affects even the smooth power spectrum. This damping term, when expanded for small $k$, has a leading-order contribution that is degenerate with the counterterm. As a result, we approximate part of the IR resummation effects by absorbing them into the counterterm. While this approach accounts for some of the IR effects, it does not fully capture the damping of the BAO signal, which remains a limitation in both pre- and post-reconstruction cases.

To account for the binning effect and ensure consistency between the theoretical predictions and the measured data, we estimate the weighted average of the power spectrum components within each \(k\)-bin. This averaged value represents the theoretical prediction corresponding to the measured data and is computed as:
\begin{equation}
P_{\ell}(k_{\text{bin}}) = \frac{\sum_{i} P_{\ell}(k_{i})}{\sum_i} 
\approx \frac{\int_{k_{\mathrm{bin,min}}}^{k_{\mathrm{bin,max}}} \mathrm{d}k \, k^{2} P_{\ell}(k)}{\int_{k_{\mathrm{bin,min}}}^{k_{\mathrm{bin,max}}} \mathrm{d}k \, k^{2}}, 
\label{eq:binning}
\end{equation} where \(k_{\mathrm{bin,min}}\) and \(k_{\mathrm{bin,max}}\) define the boundaries of the \(k\)-bin. The weighting factor \(k^{2}\) arises from the spherical volume element in Fourier space, ensuring that the theoretical model accurately reflects the binning applied to the simulated data.

\begin{figure*}
\begin{center}
\includegraphics[width=8.5cm]{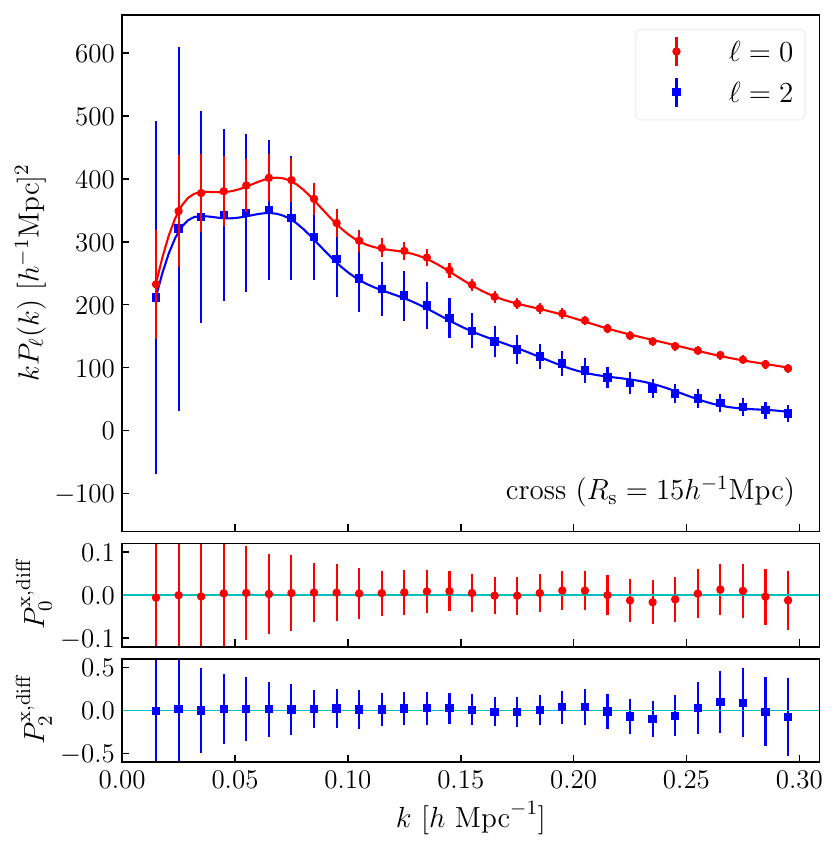}\includegraphics[width=8.5cm]{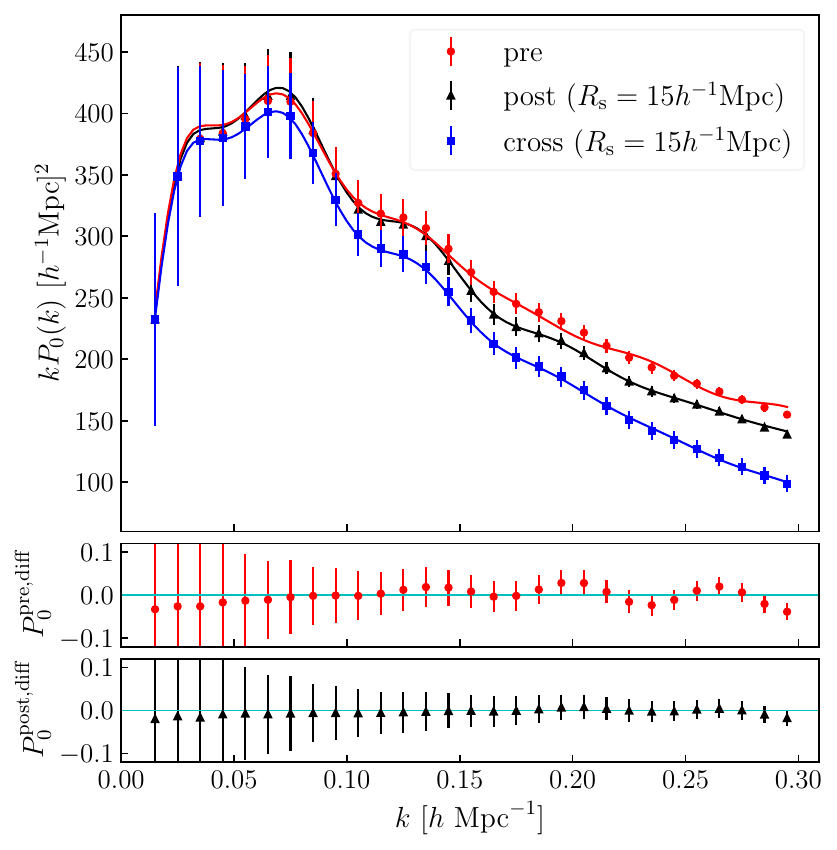}
\caption{\label{fig_plk}
Comparison of the monopole and quadrupole components for three types of matter power spectra in redshift space at \(z = 1.02\). A smoothing scale of \(R_{\mathrm{s}} = 15\, h^{-1}\mathrm{Mpc}\) is used for the reconstruction. The solid lines represent the binning-corrected theoretical predictions from the one-loop Standard Perturbation Theory (SPT), computed using the best-fit values of the linear growth rate \(f\) and the counterterm parameters \(\alpha_{\ell}\), with \(k_{\mathrm{max}} = 0.29\, h\, \mathrm{Mpc}^{-1}\). The data points for \(P_{\ell}^{\mathrm{z}(\mathrm{pre})}\) are measured from $8$ realizations of \(N\)-body simulations with a box size of \(L = 4\, h^{-1}\mathrm{Gpc}\). The data points for \(P_{\ell}^{\mathrm{z}(\mathrm{rec})}\) and \(P_{\ell}^{\mathrm{z}(\mathrm{x})}\) have been corrected according to Eq.~(\ref{eq:grid-corr}). The error bars indicate the standard deviation derived from $4,000$ realizations with a simulation box size of \(L = 500\, h^{-1}\mathrm{Mpc}\). The bottom panels in each figure display the relative difference between the simulated data and the theoretical predictions, defined as \(P_{\ell}^{\mathrm{diff}} \equiv P_{\ell}^{\mathrm{sim}} / P_{\ell}^{\mathrm{theory}} - 1\), providing a quantitative assessment of the model’s accuracy across different scales.
}
\end{center}
\end{figure*}

Figure~\ref{fig_plk} presents the monopole and quadrupole components of the cross-power spectrum (left panel) and the monopole components of the pre-reconstruction, post-reconstruction, and cross-power spectra (right panel) at redshift \(z = 1.02\). The solid lines represent the binning-corrected theoretical predictions based on the best-fit values of the linear growth rate \(f\) and the counterterm parameters, using a maximum wavenumber of \(k_{\mathrm{max}} = 0.29\, h\, \mathrm{Mpc}^{-1}\) and adopting the counterterm form described in Eq.~(\ref{eq:ct1}). The data points for \(P_{\ell}^{\mathrm{z}(\mathrm{pre})}\) are derived from larger-volume \(N\)-body simulations with a box size of \(L = 4\, h^{-1}\mathrm{Gpc}\). These are used to apply grid corrections, as outlined in Eq.~(\ref{eq:grid-corr}), to the smaller-box simulation data vectors for \(P_{\ell}^{\mathrm{z}(\mathrm{rec})}\) and \(P_{\ell}^{\mathrm{z}(\mathrm{x})}\). From the left panel, we observe that the theoretical predictions, based on the best-fit parameters, show good agreement with the \(N\)-body simulation data, particularly for the monopole component. The right panel compares the monopole components of \(P_{\ell}^{\mathrm{z}(\mathrm{rec})}\), \(P_{\ell}^{\mathrm{z}(\mathrm{x})}\), and \(P_{\ell}^{\mathrm{z}(\mathrm{pre})}\), using a reconstruction smoothing scale of \(R_{\mathrm{s}} = 15\, h^{-1}\mathrm{Mpc}\). It is evident that the Baryon Acoustic Oscillation (BAO) features on quasi-nonlinear scales are more pronounced after reconstruction, making them easier to capture with the theoretical model. The reconstruction process alters the mode coupling structure and nonlinear behavior across different scales, leading to partial de-correlation between the pre- and post-reconstruction density fields. As a result, the cross-power spectrum between the pre- and post-reconstructed density fields is suppressed compared to the auto-power spectra of the pre- and post-reconstruction fields, particularly on quasi-linear scales. This suppression becomes more significant at smaller scales due to enhanced de-correlation driven by nonlinear physical processes, which are less effectively mitigated by the reconstruction’s decoupling of coupled modes. The bottom bars in each panel of Figure~\ref{fig_plk} display the relative differences between the \(N\)-body simulation data and the theoretical predictions, defined as \(P_{\ell}^{\mathrm{diff}} \equiv P_{\ell}^{\mathrm{sim}} / P_{\ell}^{\mathrm{theory}} - 1\). Comparing these residuals for the monopole components of the three power spectrum types reveals that an appropriately calibrated post-reconstruction power spectrum model provides a better fit to the simulation data. Interestingly, the cross-power spectrum also partially inherits this improved modeling capability. By adjusting the smoothing scale, we can control the strength of the reconstruction, thereby balancing reconstruction-induced errors against improved theoretical modeling. This balance ensures the model’s applicability across a broader range of scales.

In our analysis, we consider the linear growth rate \(f\) and the counterterm parameters \(\alpha_{0}\) and \(\alpha_{2}\) as the free parameters of our theoretical model. To sample the parameter space, we employ the Markov Chain Monte Carlo (MCMC) framework implemented in {\tt Cobaya} \citep{Torrado:2020dgo}. The likelihood is evaluated using the chi-squared statistic, defined as:
\begin{equation}
\chi^{2}(\mathbf{p}) = \sum_{\ell, \ell^{\prime}}^{0,2} \sum_{i, j}^{k_{\mathrm{min}} \leq k_{i}, k_{j} \leq k_{\mathrm{max}}} 
\left[ P_{\ell}^{\mathrm{theory}}(k_{i}; \mathbf{p}) - P_{\ell}^{\mathrm{sim}}(k_{i}) \right] 
\mathbf{Cov}^{-1}_{\ell \ell^{\prime}}(k_{i}, k_{j}) 
\left[ P_{\ell^{\prime}}^{\mathrm{theory}}(k_{j}; \mathbf{p}) - P_{\ell^{\prime}}^{\mathrm{sim}}(k_{j}) \right],
\end{equation} where \(\mathbf{p} = (f, \alpha_{0}, \alpha_{2})\) represents the set of free parameters to be constrained. Here, \(P_{\ell}^{\mathrm{theory}}(k; \mathbf{p})\) and \(P_{\ell}^{\mathrm{sim}}(k)\) are the theoretical and simulated power spectrum multipoles, respectively, and \(\mathbf{Cov}^{-1}_{\ell \ell^{\prime}}(k_{i}, k_{j})\) is the inverse covariance matrix estimated from the simulations. We adopt uniform priors \(\mathcal{U}(\mathrm{min}, \mathrm{max})\) for all parameters. Specifically, we set \(\mathcal{U}(0.1, 2.0)\) for the linear growth rate \(f\) and \(\mathcal{U}(-10^{4}, 10^{4})\  [h^{-1}\mathrm{Mpc}]^2\) for the counterterm coefficients \(\alpha_{\ell}\). This choice of priors ensures a wide and unbiased exploration of the parameter space. We set 
the Gelman-Rubin statistic as \(R-1=0.001\) to ensure convergence. And the BOBYQA algorithm \citep{powell2009bobyqa,cartis2019improving,cartis2022escaping} is  utilized to minimize $\chi^{2}$ and determine the best-fit values of parameters.

\begin{figure*}
\begin{center}
\includegraphics[width=10cm]{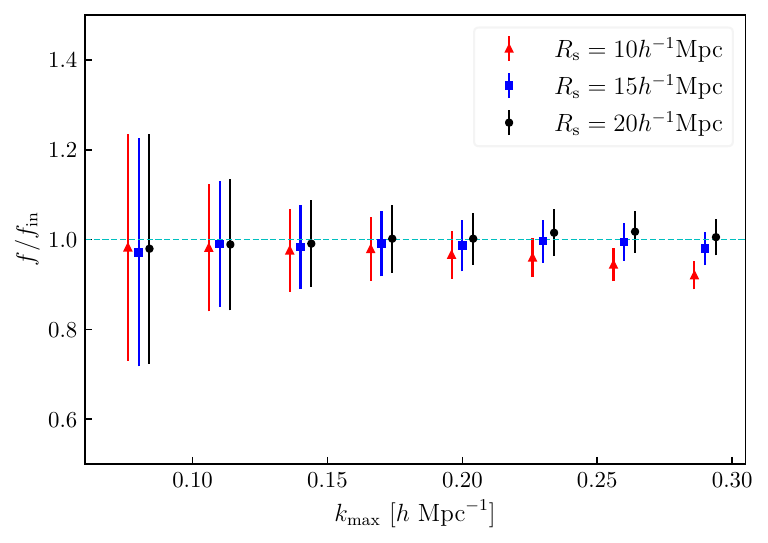}
\caption{\label{fig_bestfit}Best-fit results for the linear growth rate obtained using different smoothing scales for reconstruction. Red triangles, blue squares, and black circles represent results for \(R_{\mathrm{s}} = 10\, h^{-1}\mathrm{Mpc}\), \(15\, h^{-1}\mathrm{Mpc}\), and \(20\, h^{-1}\mathrm{Mpc}\), respectively. The error bars indicate the \(1\sigma\) uncertainties. To derive the best-fit values, we set \(k_{\mathrm{min}} = 0.01\, h\,\mathrm{Mpc}^{-1}\) and include all data vectors of the monopole (\(P_{0}\)) and quadrupole (\(P_{2}\)) up to varying \(k_{\mathrm{max}}\), from \(0.08\, h\,\mathrm{Mpc}^{-1}\) to \(0.29\, h\,\mathrm{Mpc}^{-1}\) in increments of \(0.03\, h\,\mathrm{Mpc}^{-1}\).}
\end{center}
\end{figure*}

Figure~\ref{fig_bestfit} presents the ratio of the best-fit linear growth rate \(f\) to the input value \(f_{\mathrm{in}}\), along with the corresponding 68\% confidence level (CL) uncertainties. These results account for corrections related to box-size and binning effects, and display the outcomes for different reconstruction smoothing scales. As shown in the figure, there is no significant difference among the best-fit values of \(f\) for different smoothing scales on large scales (i.e., at low \(k_{\mathrm{max}}\)). However, as smaller-scale data (larger \(k_{\mathrm{max}}\)) are included in the analysis, discrepancies become more pronounced. Specifically, when increasing \(k_{\mathrm{max}}\), the fitted values of \(f\) tend to be systematically lower for smaller smoothing scales compared to larger ones. This indicates that proper selection of the reconstruction smoothing scale is critical for obtaining unbiased measurements, especially when extending the analysis to smaller scales. Overall, the linear growth rate can be measured without significant bias using our cross-power spectrum model up to \(k_{\mathrm{max}} = 0.29\, h\, \mathrm{Mpc}^{-1}\) for smoothing scales of \(R_{\mathrm{s}} = 15\) and \(20\, h^{-1}\mathrm{Mpc}\), and up to \(k_{\mathrm{max}} = 0.20\, h\, \mathrm{Mpc}^{-1}\) for \(R_{\mathrm{s}} = 10\, h^{-1}\mathrm{Mpc}\) at redshift \(z = 1.02\). It is important to note that the simulation data used to estimate the covariance matrix are based on a box volume of \(V = (500\, h^{-1}\mathrm{Mpc})^3\), which is smaller than the volumes of typical observational surveys. Given this limitation, our model is considered robust and reliable when applied to scales up to \(k_{\mathrm{max}} = 0.20\, h\, \mathrm{Mpc}^{-1}\) for smoothing scales of \(R_{\mathrm{s}} = 15\) and \(20\, h^{-1}\mathrm{Mpc}\) at \(z = 1.02\).

\begin{figure*}
\begin{center}
\includegraphics[width=18cm]{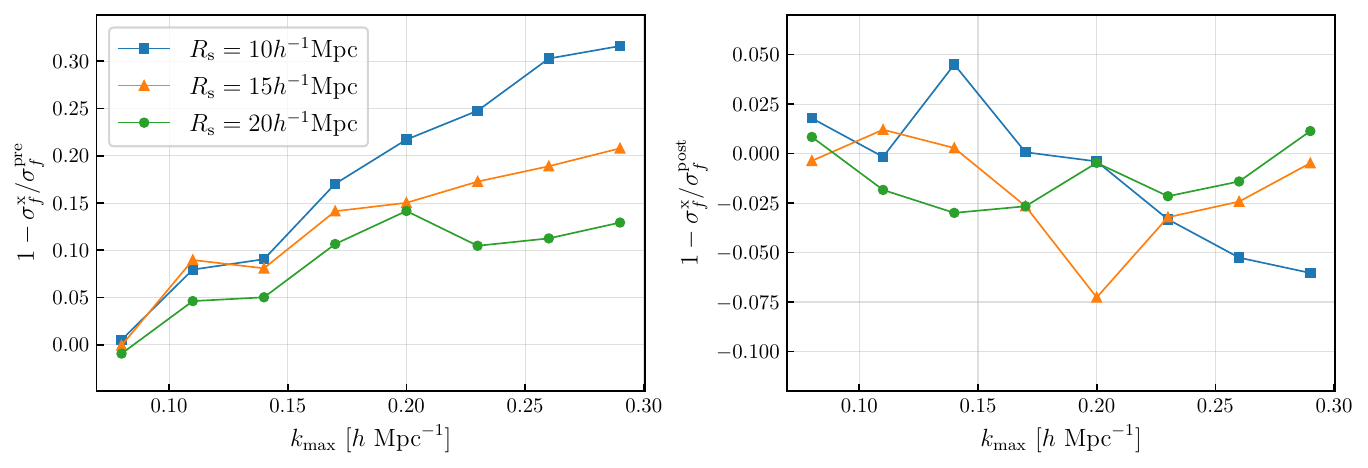}
\caption{\label{fig_sddev_chi2}
Comparison of \(\sigma_f^{\mathrm{x}}\), the measured uncertainty of the linear growth rate \(f\) from the cross-power spectrum \(P^{\mathrm{x}}\), with \(\sigma_f^{\mathrm{pre}}\) (left panel), the measured uncertainty of \(f\) from the pre-reconstruction power spectrum \(P^{\mathrm{pre}}\), and with \(\sigma_f^{\mathrm{post}}\) (right panel), the measured uncertainty of \(f\) from the post-reconstruction power spectrum \(P^{\mathrm{post}}\).
}
\end{center}
\end{figure*}

Figure~\ref{fig_sddev_chi2} compares the uncertainties in the linear growth rate \(f\) obtained from different power spectra. The left panel shows the comparison between \(\sigma_f^{\mathrm{x}}\), the uncertainty derived from the cross-power spectrum \(P^{\mathrm{x}}\), and \(\sigma_f^{\mathrm{pre}}\), the uncertainty obtained from the pre-reconstruction power spectrum \(P^{\mathrm{pre}}\). The right panel compares \(\sigma_f^{\mathrm{x}}\) with \(\sigma_f^{\mathrm{post}}\), the uncertainty measured from the post-reconstruction power spectrum \(P^{\mathrm{post}}\). As shown in the figure, \(\sigma_f^{\mathrm{x}}\) exhibits intermediate behavior, generally smaller than \(\sigma_f^{\mathrm{pre}}\) but larger than \(\sigma_f^{\mathrm{post}}\) across most scales. This indicates that the cross-power spectrum provides complementary information, effectively reducing the uncertainty in \(f\) compared to pre-reconstruction measurements while not fully reaching the precision achieved with post-reconstruction data. The results highlight the utility of incorporating cross-correlations in enhancing the constraints on the linear growth rate, particularly in the context of cosmological parameter estimation.

\begin{figure*}
\begin{center}
\includegraphics[width=12cm]{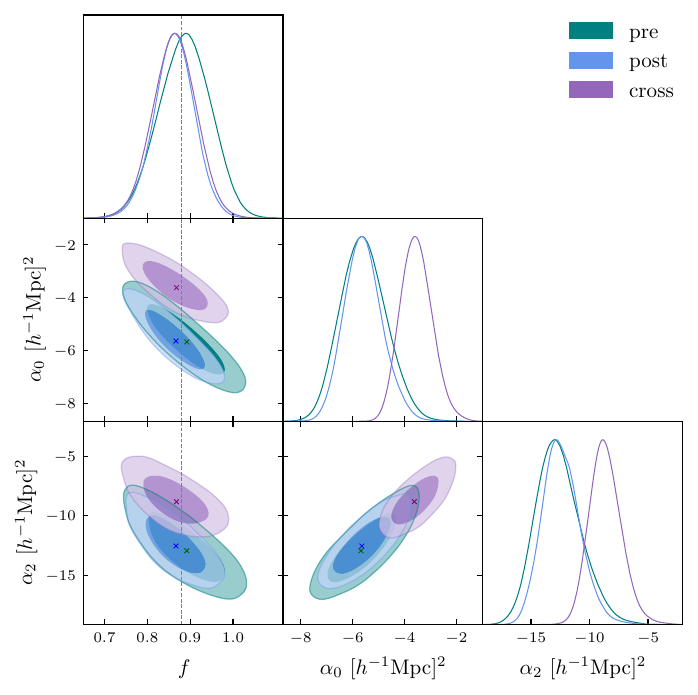}
\caption{\label{fig_contour}
Comparison of the constraints on the linear growth rate \(f\) and the counterterm parameters \(\alpha_0\) and \(\alpha_2\) derived from the pre-reconstruction (green), post-reconstruction (blue), and cross (purple) matter power spectra. The same prior settings are applied to each model. The data vectors for the monopole (\(P_0\)) and quadrupole (\(P_2\)) are used within the wavenumber range \(k_{\mathrm{min}} = 0.01\, h\, \mathrm{Mpc}^{-1}\) to \(k_{\mathrm{max}} = 0.20\, h\, \mathrm{Mpc}^{-1}\), with a reconstruction smoothing scale of \(R_{\mathrm{s}} = 15\, h^{-1}\mathrm{Mpc}\). The best-fit parameter values are indicated by the x-shaped markers, while the vertical dashed line denotes the fiducial value of the linear growth rate, \(f_{\mathrm{in}} = 0.8796\).
}
\end{center}
\end{figure*}

Using the {\tt GetDist} software \citep{Lewis:2019xzd}, Figure~\ref{fig_contour} presents the 68\% and 95\% confidence level contours, along with the one-dimensional posterior distributions, for the linear growth rate \(f\) and the counterterm parameters \(\alpha_0\) and \(\alpha_2\), derived from three types of power spectra. For this comparison, we adopt a reconstruction smoothing scale of \(R_{\mathrm{s}} = 15\, h^{-1}\mathrm{Mpc}\) and a maximum wavenumber of \(k_{\mathrm{max}} = 0.20\, h\, \mathrm{Mpc}^{-1}\). The figure shows that the redshift-space distortion (RSD) parameter \(f\) is more tightly constrained after density field reconstruction, as expected. The constraint on \(f\) from the cross-power spectrum \(P^{\mathrm{x}}\) is comparable to that from the post-reconstruction power spectrum \(P^{\mathrm{post}}\), with a relative improvement of approximately \(1 - \sigma_f^{\mathrm{x}} / \sigma_f^{\mathrm{post}} = -7.3\%\). Although the cross-power spectrum does not outperform the post-reconstruction power spectrum in constraining cosmological parameters, we find that our model for \(P^{\mathrm{x}}\) yields tighter constraints on the counterterm parameters compared to both the pre- and post-reconstruction cases. Counterterm parameters encapsulate the contributions from small-scale (UV) physics to the large-scale modes of interest. Interestingly, the absolute values of the counterterm parameters for the pre- and post-reconstruction cases are similar, while those for the cross-power spectrum are notably smaller. The differences in the counterterms between pre-, post-, and cross-power spectra could be partially explained by the lack of IR cancellation in the cross-spectrum. Since IR resummation is not explicitly performed, the damping effect modifies the amplitude of the power spectrum in a way that can lead to a systematic shift in the counterterm parameters. This may contribute to the observed difference in $\alpha_0$ between the pre/post-reconstruction spectra and the cross-spectrum, as illustrated in Fig.~\ref{fig_contour}. In the absence of IR cancellation, residual effects alter the shape of the cross-power spectrum, particularly at small scales. These effects can mimic or interfere with the counterterm contributions, complicating the interpretation of the parameter differences. A more detailed investigation incorporating IR resummation explicitly would be needed to disentangle these contributions more rigorously. Future work incorporating IR-resummed theoretical models for reconstructed cases \citep{Sugiyama:2024eye, Sugiyama:2024qsw} will help clarify these effects and improve the modeling of the cross power spectrum.

\section{Discussions and conclusion} 
\label{sec:conclusion}

The nonlinear evolution of cosmic matter density fluctuations induces non-Gaussian features in the nonlinear density field \(\delta\), which cannot be fully captured by two-point statistics such as the power spectrum. However, Baryon Acoustic Oscillation (BAO) reconstruction techniques, which act as a nonlinear inverse transformation of \(\delta\), aim to partially reverse the nonlinear effects that dampen and shift the BAO signal \citep{Crocce:2007dt, Seo:2008yx}. The standard BAO reconstruction method, based on the simple yet robust Zel’dovich approximation to compute the shift field through negative displacements, provides an effective approach to linearizing the nonlinear density field. As a result, the reconstructed density field \(\delta^{(\mathrm{rec})}\) appears more linear, containing a larger fraction of accessible Gaussian information compared to the pre-reconstructed density field \(\delta\). Consequently, some of the higher-order statistical information embedded in \(\delta\) can be partially recovered through the two-point statistics of \(\delta^{(\mathrm{rec})}\). While the reconstruction process alters the relationship between cosmological parameters and the density fields, it does not inherently change the parameter values themselves—though if such changes occurred, they would require precise modeling. This reconstruction-induced modification of statistical properties enables the breaking of parameter degeneracies through a joint analysis of the pre-reconstruction power spectrum (\(P^{\mathrm{pre}}\)), post-reconstruction power spectrum (\(P^{\mathrm{post}}\)), and the cross-power spectrum (\(P^{\mathrm{x}}\)) {\citep{Wang:2022nlx,prepostEmu,Zhang:2025urx}}. This highlights the complementary role of \(P^{\mathrm{x}}\) in tightening cosmological parameter constraints. In this context, we present a theoretical model for the cross-power spectrum of pre- and post-reconstructed matter density fields in redshift space, developed within the framework of perturbation theory.
 
The one-loop terms of \(P^{\mathrm{z}(\mathrm{x})}\) can be expressed in the same form as those for \(P^{\mathrm{z}}\) and \(P^{\mathrm{z}(\mathrm{rec})}\), with redefined perturbation kernels \(F_{22}^{\mathrm{z}(\mathrm{x})}\) and \(F_{13}^{\mathrm{z}(\mathrm{x})}\). Using a reconstruction smoothing scale of \(R_{\mathrm{s}} = 15\, h^{-1}\mathrm{Mpc}\) as an example, we compare the monopole and quadrupole components of \(F_{22}^{\mathrm{z}(\mathrm{x})}(k, \mu)\) and \(F_{13}^{\mathrm{z}(\mathrm{x})}(k, \mu)\) with those of the pre- and post-reconstruction cases. The \(P_{22}\) term in the one-loop Standard Perturbation Theory (SPT) describes the contribution of mode coupling to the nonlinear power spectrum, while the \(P_{13}\) term reflects the suppression of the power spectrum due to nonlinear structure evolution. Similar to the post-reconstruction case, we observe a significant reduction in the amplitudes of \(F_{22,\ell=0,2}^{\mathrm{z}(\mathrm{x})}\), indicating a weakened influence of mode coupling as a result of the reconstruction. Additionally, the negative amplitudes of \(F_{13,\ell=0,2}^{\mathrm{z}(\mathrm{x})}\) are also reduced compared to the pre-reconstruction case, reflecting the diminished impact of nonlinear growth. However, the absolute reductions in the amplitudes of \(F_{22,\ell=0,2}^{\mathrm{z}(\mathrm{x})}\) and \(F_{13,\ell=0,2}^{\mathrm{z}(\mathrm{x})}\) are generally smaller than those observed in the post-reconstruction case. On large scales, the net one-loop contribution to \(P^{\mathrm{z}(\mathrm{x})}\) is predominantly negative for both the monopole and quadrupole components. This behavior arises from the lack of substantial cancellation between \(P_{22}^{\mathrm{z}(\mathrm{x})}\) and \(P_{13}^{\mathrm{z}(\mathrm{x})}\), which contrasts with the partial cancellations seen in the pre- and post-reconstruction cases. The larger negative amplitudes of the net one-loop contributions in \(P^{\mathrm{z}(\mathrm{x})}\), compared to \(P^{\mathrm{z}}\) and \(P^{\mathrm{z}(\mathrm{rec})}\), suggest a stronger decorrelation between the pre- and post-reconstructed density fields. This decorrelation is a key feature of the cross-power spectrum, reflecting the distinct nonlinear evolution of the density fields before and after reconstruction. Moreover, since the cross covariance between the pre- and post-reconstruction power spectra is mainly determined by $P^{\mathrm{x}}$ \citep{Zhao:2024xit}, the decorrelation feature indicates the existence of complementary information between these two observables.

By analyzing the infrared (IR) behavior of the one-loop integrals of \(P^{\mathrm{z}(\mathrm{x})}\), we find that the contributions from the estimated shift field persist even after summing the IR limits of \(P_{22}^{\mathrm{z}(\mathrm{x})}\) and \(P_{13}^{\mathrm{z}(\mathrm{x})}\). This results in a net negative power contribution to the one-loop components of \(P^{\mathrm{z}(\mathrm{x})}\), which leads to the observed decorrelation between the pre-reconstructed density field \(\delta\) and the post-reconstructed density field \(\delta^{(\mathrm{rec})}\). Furthermore, we investigate the ultraviolet (UV) limits of the one-loop terms of \(P^{\mathrm{x}}\) in real space. We find that the leading-order contributions from the reconstruction components of \(P_{22}^{\mathrm{x}}\) and \(P_{13}^{\mathrm{x}}\) exhibit different \(k\)-dependencies compared to their pre-reconstruction counterparts. However, these contributions are significantly suppressed by the exponential decay factors associated with the smoothing kernel \(W\). As a result, the dominant UV behavior of \(P_{22}^{\mathrm{x}}\) and \(P_{13}^{\mathrm{x}}\) remains consistent with that of the pre-reconstruction case. This finding supports the incorporation of the lowest-order Effective Field Theory (EFT) counterterms into our model to effectively parameterize the residual effects of UV physics.

We validate the model of \(P^{\mathrm{x}}\) by fitting the linear growth rate \(f\) and the counterterm parameters \(\alpha_0\) and \(\alpha_2\) to the measured monopole and quadrupole moments, \(P_{\ell=0,2}^{\mathrm{z}(\mathrm{x})}\), derived from \(N\)-body simulations at \(z = 1.02\). For appropriate smoothing scales, such as \(R_{\mathrm{s}} = 15\, h^{-1}\mathrm{Mpc}\) and \(20\, h^{-1}\mathrm{Mpc}\), the recovered values of \(f\) are consistent with the fiducial values across different \(k\)-ranges. However, significant deviations from the fiducial value occur when using smaller smoothing scales and incorporating data from higher wavenumbers, a trend also observed in the post-reconstruction case. When comparing the constraining power of \(P^{\mathrm{z}(\mathrm{x})}\) on \(f\) to the pre-reconstruction case, we find an improvement of approximately \(14\%\) to \(22\%\) for \(k \leq 0.20\, h\, \mathrm{Mpc}^{-1}\). 
{This demonstrates that the cross-power spectrum can provide better constraints on the linear growth rate compared to the pre-reconstructed case. Our analysis shows that when each of the three power spectra is analyzed individually, the post-reconstruction power spectrum provides the strongest constraints on the linear growth rate.
Current galaxy survey analyses primarily use reconstruction for BAO constraints \citep[e.g.,][]{eBOSS:2020fvk,Gil-Marin:2022hnv}, but studies based on numerical simulations and perturbation theory have demonstrated that the post-reconstruction power spectrum, with a properly chosen smoothing scale, can also reliably improve RSD measurements \citep[e.g.,][]{Hikage:2019ihj,Hikage:2020fte,prepostEmu}. When fixing BAO parameters, our results show that the post-reconstruction power spectrum generally provide the strongest constraints on  \(f\) among the three types of power spectra. However,} 
we find that the counter-term parameters of \(P^{\mathrm{z}(\mathrm{x})}\) are better constrained than those of both the pre- and post-reconstruction power spectra. 
{Note that these results are obtained under the assumption of fixed cosmological parameters and only investigates the impact of the cross-power spectrum on the constraints of the linear growth rate and counter-term parameters. In more general cases, when cosmological parameters are allowed to vary simultaneously, counter-terms tend to become more weakly constrained, and the cross-power spectrum’s constraints on counter-terms may change. This issue can be explored further in future studies. Although the counter-terms in different power spectrum models typically represent different physical aspects, the three reconstruction-related power spectra all originate from the pre-reconstruction density field. As a result, there is likely a relationship between their counter-terms that merits further investigation. This relationship is crucial for the joint analysis of the three power spectra, as it could help reduce the number of free parameters involved in the analysis.} If the relationship between the counterterms of the pre-, post-reconstruction, and cross-power spectra can be accurately established, the improved constraints on these counterterms could enhance joint analyses involving all three power spectra. The comparison between the model predictions, using the best-fit parameter values, and the simulation measurements shows that the \(P^{\mathrm{z}(\mathrm{x})}\) model provides a good fit to the simulated data up to \(k_{\mathrm{max}} = 0.29\, h\, \mathrm{Mpc}^{-1}\) with \(R_{\mathrm{s}} = 15\, h^{-1}\mathrm{Mpc}\). Moreover, the BAO features are better captured after reconstruction, suggesting that IR resummation may have a less significant impact on \(P^{\mathrm{z}(\mathrm{rec})}\) and \(P^{\mathrm{z}(\mathrm{x})}\) compared to the pre-reconstruction case. However, for a consistent analysis of \(P^{\mathrm{z}(\mathrm{pre})}\), \(P^{\mathrm{z}(\mathrm{rec})}\), and \(P^{\mathrm{z}(\mathrm{x})}\), it is necessary to develop IR-resummed theoretical models for the reconstructed cases, as discussed in \citep{Sugiyama:2024eye, Sugiyama:2024qsw}.

In future work, we plan to apply our model to the joint analysis of the pre-, post-reconstruction, and cross-power spectra. Additionally, observational effects such as the Alcock–Paczynski (AP) effect, galaxy bias, window functions, and other systematic factors will need to be incorporated for future applications in the analysis of galaxy survey data.

\begin{acknowledgments}

We thank Chiaki Hikage for contributions in the early stage of this work. WZ, RZ, XM, YW, and GBZ are supported by the National Natural Science Foundation of China (NSFC) under Grant 11925303. KK is supported by the Science and Technology Facilities Council (STFC) under Grant ST/W001225/1. RT is supported by JSPS KAKENHI grant Nos. JP22H00130 and JP20H05855.YW further acknowledges support from the National Key R\&D Program of China No. (2022YFF0503404, 2023YFA1607800, 2023YFA1607803), NSFC Grants 12273048 and 12422301, the CAS Project for Young Scientists in Basic Research (No. YSBR-092), and the Youth Innovation Promotion Association of CAS. GBZ also acknowledges support from the CAS Project for Young Scientists in Basic Research (No. YSBR-092), the China Manned Space Project, and the New Cornerstone Science Foundation through the XPLORER Prize.

\end{acknowledgments}

\newpage

\appendix
\section{Derivation of the Cross Power Spectrum}\label{de_px}

Starting from the definition of the power spectrum, we derive the
expression for the cross power spectrum using a perturbation series
for the pre- and post-reconstructed matter density field. Substituting
Eq.~(\ref{eq:dkz}) and Eq.~(\ref{eq:dkzr}) into the expanded form
of the left-hand side of Eq.~(\ref{eq:pzxk}), we obtain:
\begin{equation}
\left\langle \tilde{\delta}_{\mathbf{k}}^{\mathrm{z}}\tilde{\delta}_{\mathbf{k}^{\prime}}^{\mathrm{z}(\mathrm{rec})}\right\rangle =\sum_{m,n=1}^{\infty}\left\langle \tilde{\delta}_{\mathbf{k}}^{\mathrm{z}(m)}\tilde{\delta}_{\mathbf{k}^{\prime}}^{\mathrm{z}(\mathrm{rec})(n)}\right\rangle ,
\end{equation}
where
\begin{eqnarray}
\left\langle \tilde{\delta}_{\mathbf{k}}^{\mathrm{z}(m)}\tilde{\delta}_{\mathbf{k}^{\prime}}^{\mathrm{z}(\mathrm{rec})(n)}\right\rangle  & = & D^{m+n}(z)\int\frac{\mathrm{d}\mathbf{p}_{1}\cdots\mathrm{d}\mathbf{p}_{m}\mathrm{d}\mathbf{p}_{m+1}\cdots\mathrm{d}\mathbf{p}_{m+n}}{(2\pi)^{3(m+n)-6}}\,\delta_{\mathrm{D}}\left(\sum_{i=1}^{m}\mathbf{p}_{i}-\mathbf{k}\right)\delta_{\mathrm{D}}\left(\sum_{j=1}^{n}\mathbf{p}_{m+j}-\mathbf{k}^{\prime}\right)\nonumber \\
 &  & \times F_{m}^{\mathrm{z}}\left(\mathbf{p}_{1},...,\mathbf{p}_{m}\right)F_{n}^{\mathrm{z}(\mathrm{rec})}\left(\mathbf{p}_{m+1},...,\mathbf{p}_{m+n}\right)\left\langle \tilde{\delta}_{\mathbf{p}_{1}}^{\mathrm{L}}\cdots\tilde{\delta}_{\mathbf{p}_{m}}^{\mathrm{L}}\tilde{\delta}_{\mathbf{p}_{m+1}}^{\mathrm{L}}\cdots\tilde{\delta}_{\mathbf{p}_{m+n}}^{\mathrm{L}}\right\rangle .\label{eq:ddr}
\end{eqnarray}
This equation expresses the cross-correlator between the pre- and
post-reconstructed matter density perturbations. Note that the cases
where $m\neq n$ differ slightly from the auto-correlation as they
are not commutatively symmetric. 

Assuming that the linear overdensity $\delta^{\mathrm{L}}$ follows
a Gaussian distribution, Wick’s theorem allows us to compute the even-point
correlation of the linear overdensity in Eq.~(\ref{eq:ddr}) as \citep{Bernardeau:2001qr}:
\begin{equation}
\left\langle \tilde{\delta}_{\mathbf{p}_{1}}^{\mathrm{L}}\cdots\tilde{\delta}_{\mathbf{p}_{m+n}}^{\mathrm{L}}\right\rangle =\sum_{\text{all pair associations}}\prod_{(i,j)\in\text{pairs}}\left\langle \tilde{\delta}_{\mathbf{p}_{i}}^{\mathrm{L}}\tilde{\delta}_{\mathbf{p}_{j}}^{\mathrm{L}}\right\rangle ,\quad\text{if }(m+n)\text{ is even}.
\end{equation}
Using the definition of the linear power spectrum,
\begin{equation}
\left\langle \tilde{\delta}_{\mathbf{k}}^{\mathrm{L}}\tilde{\delta}_{\mathbf{k}^{\prime}}^{\mathrm{L}}\right\rangle \equiv(2\pi)^{3}\delta_{\mathrm{D}}(\mathbf{k}+\mathbf{k}^{\prime})P_{\mathrm{L}}(k),
\end{equation}
and the properties of the Dirac delta function $\delta_{\mathrm{D}}$,
we can perform the momentum integration in Eq.~(\ref{eq:ddr}), which
can also be done systematically using the diagrammatic method. The
nonzero correlators contributing to the power spectrum up to one-loop
order are:
\begin{equation}
\left\langle \tilde{\delta}_{\mathbf{k}}^{\mathrm{z}(1)}\tilde{\delta}_{\mathbf{k}^{\prime}}^{\mathrm{z}(\mathrm{rec})(1)}\right\rangle =(2\pi)^{3}\delta_{\mathrm{D}}(\mathbf{k}+\mathbf{k}^{\prime})D^{2}(z)F_{1}^{\mathrm{z}}(\mathbf{k})F_{1}^{\mathrm{z}(\mathrm{rec})}(\mathbf{k})P_{\mathrm{L}}(k),
\end{equation}
\begin{equation}
\left\langle \tilde{\delta}_{\mathbf{k}}^{\mathrm{z}(2)}\tilde{\delta}_{\mathbf{k}^{\prime}}^{\mathrm{z}(\mathrm{rec})(2)}\right\rangle =(2\pi)^{3}\delta_{\mathrm{D}}(\mathbf{k}+\mathbf{k}^{\prime})D^{4}(z)\times2\int\frac{\mathrm{d}\mathbf{p}}{(2\pi)^{3}}F_{2}^{\mathrm{z}}(\mathbf{k}-\mathbf{p},\mathbf{p})F_{2}^{\mathrm{z}(\mathrm{rec})}(\mathbf{k}-\mathbf{p},\mathbf{p})P_{\mathrm{L}}(|\mathbf{k}-\mathbf{p}|)P_{\mathrm{L}}(p),
\end{equation}
\begin{equation}
\left\langle \tilde{\delta}_{\mathbf{k}}^{\mathrm{z}(1)}\tilde{\delta}_{\mathbf{k}^{\prime}}^{\mathrm{z}(\mathrm{rec})(3)}\right\rangle =(2\pi)^{3}\delta_{\mathrm{D}}(\mathbf{k}+\mathbf{k}^{\prime})D^{4}(z)\times3F_{1}^{\mathrm{z}}(\mathbf{k})P_{\mathrm{L}}(k)\int\frac{\mathrm{d}\mathbf{p}}{(2\pi)^{3}}F_{3}^{\mathrm{z}(\mathrm{rec})}(\mathbf{k},\mathbf{p},-\mathbf{p})P_{\mathrm{L}}(p),
\end{equation}
\begin{equation}
\left\langle \tilde{\delta}_{\mathbf{k}}^{\mathrm{z}(3)}\tilde{\delta}_{\mathbf{k}^{\prime}}^{\mathrm{z}(\mathrm{rec})(1)}\right\rangle =(2\pi)^{3}\delta_{\mathrm{D}}(\mathbf{k}+\mathbf{k}^{\prime})D^{4}(z)\times3F_{1}^{\mathrm{z}(\mathrm{rec})}(\mathbf{k})P_{\mathrm{L}}(k)\int\frac{\mathrm{d}\mathbf{p}}{(2\pi)^{3}}F_{3}^{\mathrm{z}}(\mathbf{k},\mathbf{p},-\mathbf{p})P_{\mathrm{L}}(p).
\end{equation}
Since the first-order perturbation kernels for the pre- and post-reconstructed
density fields in redshift space are the same, we have:
\begin{equation}
F_{1}^{\mathrm{z}}(\mathbf{k})=F_{1}^{\mathrm{z}(\mathrm{rec})}(\mathbf{k})=1+f\mu^{2},
\end{equation}
where $\mu=(\mathbf{k}\cdot\hat{\mathbf{z}})/k$. Thus, the one-loop
corrections $P_{22}^{\mathrm{z}(\mathrm{x})}$ and $P_{13}^{\mathrm{z}(\mathrm{x})}$
are given by Eqs.~(\ref{eq:p22}) and (\ref{eq:p13}) for formal consistency
with the auto-correlated power spectrum.

\section{Formulae for the 1-Loop Power Spectrum}

The one-loop terms $P_{22}^{\mathrm{z}(\mathrm{x})}$ and $P_{13}^{\mathrm{z}(\mathrm{x})}$
can be computed in a spherical coordinate system $(r,x,\phi)$, where
$\mathbf{k}\cdot\mathbf{p}=kpx$, $p=kr$, and $\phi$ denotes the
azimuthal angle of $\mathbf{p}$. Due to rotational symmetry, the
integration over $\phi$ can be performed analytically. The resulting
one-loop terms of the cross power spectrum are given by:
\begin{equation}
P_{22}^{\mathrm{z}(\mathrm{x})}(k,\mu)=\sum_{n,m}\mu^{2n}f^{m}\frac{k^{3}}{4\pi^{2}}\int_{0}^{\infty}\mathrm{d}r\,P_{\mathrm{L}}(kr)\int_{-1}^{1}\mathrm{d}x\,P_{\mathrm{L}}\left(k\sqrt{1+r^{2}-2rx}\right)\frac{A_{nm}^{\mathrm{x}}(r,x)}{\left(1+r^{2}-2rx\right)^{2}},
\end{equation}
\begin{equation}
P_{13}^{\mathrm{z}(\mathrm{x})}(k,\mu)=(1+f\mu^{2})P_{\mathrm{L}}(k)\sum_{n,m}\mu^{2n}f^{m}\frac{k^{3}}{4\pi^{2}}\int_{0}^{\infty}\mathrm{d}r\,P_{\mathrm{L}}(kr)\int_{-1}^{1}\mathrm{d}x\,B_{nm}^{\mathrm{x}}(r,x).
\end{equation}
The explicit expressions for the non-vanishing coefficients $A_{nm}^{\mathrm{x}}$
and $B_{nm}^{\mathrm{x}}$ are summarized below. Here, we introduce
the notation: $W_{p}\equiv W(|\mathbf{p}|)$, $W_{*}\equiv W(|\mathbf{k}-\mathbf{p}|)$
for convenience.
\begin{eqnarray}
A_{00}^{\mathrm{x}} & = & \frac{1}{98}\left[r\left(10x^{2}-3\right)-7x\right]\left[7xW_{p}\left(r^{2}-2rx+1\right)-7rW_{*}\left(rx-1\right)+10rx^{2}-3r-7x\right]\\
A_{01}^{\mathrm{x}} & = & -\frac{x^{2}-1}{28\left(r^{2}-2rx+1\right)}\left(r^{3}\left(6-20x^{2}\right)+4r^{2}x\left(5x^{2}+2\right)+r\left(3-24x^{2}\right)+7x\right)\left(xW_{p}\left(-r^{2}+2rx-1\right)\right.\nonumber \\
 &  & \left.+rW_{*}\left(rx-1\right)\right)\\
A_{02}^{\mathrm{x}} & = & \frac{3r^{2}\left(x^{2}-1\right)^{2}\left[r\left(10x^{2}-3\right)-7x\right]\left[xW_{p}\left(r^{2}-2rx+1\right)+rW_{*}\left(1-rx\right)\right]}{112\left(r^{2}-2rx+1\right)}\\
A_{11}^{\mathrm{x}} & = & \frac{r\left(10x^{2}-3\right)-7x}{196\left(r^{2}-2rx+1\right)}\left[7xW_{p}\left(r^{4}\left(6x^{2}+2\right)-2r^{3}x\left(9x^{2}+7\right)+r^{2}\left(12x^{4}+29x^{2}+7\right)-4rx\left(3x^{2}+5\right)\right.\right.\nonumber \\
 &  & \left.+3x^{2}+5\right)-7rW_{*}\left(2r^{3}\left(3x^{3}+x\right)-2r^{2}\left(3x^{4}+8x^{2}+1\right)+3rx\left(3x^{2}+5\right)-3x^{2}-5\right)\nonumber \\
 &  & \left.+8\left(r^{3}\left(10x^{2}-3\right)-r^{2}\left(20x^{3}+x\right)+3r\left(8x^{2}-1\right)-7x\right)\right]\\
A_{12}^{\mathrm{x}} & = & \frac{x^{2}-1}{56\left(r^{2}-2rx+1\right)}\left[xW_{p}\left(r^{5}\left(-150x^{4}-117x^{2}+1\right)+r^{4}x\left(300x^{4}+651x^{2}+113\right)-4r\left(113x^{2}+6\right)\right.\right.\nonumber \\
 &  & \left.-r^{3}\left(984x^{4}+659x^{2}+23\right)+r^{2}x\left(1041x^{2}+233\right)+70x\right)+rW_{*}\left(r^{4}x\left(150x^{4}+117x^{2}-1\right)\right.\nonumber \\
 &  & \left.+r^{3}\left(-567x^{4}-232x^{2}+1\right)+r^{2}x\left(729x^{2}+139\right)-2r\left(191x^{2}+12\right)+70x\right)\nonumber \\
 &  & \left.-2\left(6r^{4}\left(8x^{2}-1\right)-6r^{3}x\left(16x^{2}+5\right)+r^{2}\left(132x^{2}+1\right)-56rx+7\right)\right]\\
A_{13}^{\mathrm{x}} & = & \frac{3r\left(x^{2}-1\right)^{2}\left(r^{2}\left(76x^{2}+8\right)-84rx+21\right)\left(xW_{p}\left(r^{2}-2rx+1\right)+rW_{*}(1-rx)\right)}{112\left(r^{2}-2rx+1\right)}\\
A_{14}^{\mathrm{x}} & = & -\frac{5r^{3}\left(x^{2}-1\right)^{3}\left(xW_{p}\left(r^{2}-2rx+1\right)+rW_{*}(1-rx)\right)}{32\left(r^{2}-2rx+1\right)}\\
A_{22}^{\mathrm{x}} & = & \frac{1}{784\left(r^{2}-2rx+1\right)}\left[-7xW_{p}\left(r^{5}\left(-\left(350x^{6}+299x^{4}-324x^{2}+11\right)\right)+r^{4}x\left(700x^{6}+1947x^{4}\right.\right.\right.\nonumber \\
 &  & \left.-698x^{2}-269\right)+r^{3}\left(-3048x^{6}-1399x^{4}+1002x^{2}+85\right)+r^{2}x\left(3749x^{4}+66x^{2}-455\right)\nonumber \\
 &  & \left.+r\left(-1816x^{4}+40x^{2}+96\right)+28\left(11x^{3}+x\right)\right)\nonumber \\
 &  & -7rW_{*}\left(r^{4}x\left(350x^{6}+299x^{4}-324x^{2}+11\right)-r^{3}\left(1699x^{6}+249x^{4}-615x^{2}+11\right)\right.\nonumber \\
 &  & \left.+r^{2}x\left(2549x^{4}-146x^{2}-387\right)+r\left(-1508x^{4}+68x^{2}+96\right)+28\left(11x^{3}+x\right)\right)\nonumber \\
 &  & +4\left(6r^{4}\left(236x^{4}-157x^{2}+19\right)-6r^{3}x\left(472x^{4}+x^{2}-81\right)+r^{2}\left(5196x^{4}-1733x^{2}+65\right)\right.\nonumber \\
 &  & \left.\left.+r\left(812x-3164x^{3}\right)+637x^{2}-49\right)\right]\\
A_{23}^{\mathrm{x}} & = & \frac{x^{2}-1}{56\left(r^{2}-2rx+1\right)}\left[xW_{p}\left(r^{5}\left(-580x^{4}+108x^{2}+24\right)+4r^{4}x\left(290x^{4}+209x^{2}-51\right)\right.\right.\nonumber \\
 &  & \left.+r^{3}\left(-2684x^{4}-159x^{2}+57\right)+2r^{2}x\left(1105x^{2}-62\right)+r\left(33-775x^{2}\right)+98x\right)\nonumber \\
 &  & +rW_{*}\left(4r^{4}x\left(145x^{4}-27x^{2}-6\right)-24r^{3}\left(68x^{4}-11x^{2}-1\right)+7r^{2}x\left(233x^{2}-27\right)+r\left(33-677x^{2}\right)\right.\nonumber \\
 &  & \left.+98x\right)\left.-4\left(6r^{4}\left(8x^{2}-1\right)-6r^{3}x\left(16x^{2}+5\right)+r^{2}\left(132x^{2}+1\right)-56rx+7\right)\right]
\end{eqnarray}
\begin{eqnarray}
A_{24}^{\mathrm{x}} & = & \frac{3r\left(x^{2}-1\right)^{2}}{32\left(r^{2}-2rx+1\right)}\left[xW_{p}\left(5r^{4}\left(7x^{2}-1\right)-10r^{3}x\left(7x^{2}+2\right)+r^{2}\left(95x^{2}+1\right)-42rx+6\right)\right.\nonumber \\
 &  & \left.+r\left(2\left(r^{2}-2rx+1\right)+W_{*}\left(r^{3}\left(5x-35x^{3}\right)+r^{2}\left(65x^{2}-5\right)-36rx+6\right)\right)\right]\\
A_{33}^{\mathrm{x}} & = & \frac{1}{112\left(r^{2}-2rx+1\right)}\left[xW_{p}\left(4r^{5}\left(273x^{6}-220x^{4}-3x^{2}+6\right)-4r^{4}x\left(546x^{6}+159x^{4}-452x^{2}+27\right)\right.\right.\nonumber \\
 &  & +r^{3}\left(5884x^{6}-2801x^{4}-870x^{2}+27\right)+r^{2}\left(-5690x^{5}+3376x^{3}+74x\right)\nonumber \\
 &  & \left.+r\left(2375x^{4}-1258x^{2}+3\right)+28x\left(5-13x^{2}\right)\right)+rW_{*}\left(-4r^{4}x\left(273x^{6}-220x^{4}-3x^{2}+6\right)\right.\nonumber \\
 &  & +8r^{3}\left(436x^{6}-333x^{4}+6x^{2}+3\right)+r^{2}\left(-4043x^{5}+2762x^{3}-63x\right)\nonumber \\
 &  & \left.+r\left(2011x^{4}-1118x^{2}+3\right)+28x\left(5-13x^{2}\right)\right)\nonumber \\
 &  & +8\left(r^{4}\left(88x^{4}-66x^{2}+6\right)+r^{3}\left(-176x^{5}+22x^{3}+42x\right)\right.\nonumber \\
 &  & +r^{2}\left(308x^{4}-139x^{2}-1\right)\nonumber \\
 &  & \left.\left.+r\left(68x-180x^{3}\right)+35x^{2}-7\right)\right]\\
A_{34}^{\mathrm{x}} &  & =\frac{x^{2}-1}{32\left(r^{2}-2rx+1\right)}\left[-3xW_{p}\left(5r^{5}\left(21x^{4}-14x^{2}+1\right)+r^{4}\left(50x-210x^{5}\right)+r^{3}\left(385x^{4}-130x^{2}-7\right)\right.\right.\nonumber \\
 &  & \left.+r^{2}\left(76x-260x^{3}\right)+4r\left(19x^{2}-3\right)-8x\right)+3rW_{*}\left(5r^{4}\left(21x^{5}-14x^{3}+x\right)\right.\nonumber \\
 &  & \left.-5r^{3}\left(49x^{4}-26x^{2}+1\right)+8r^{2}x\left(25x^{2}-9\right)+r\left(12-68x^{2}\right)+8x\right)\nonumber \\
 &  & \left.+4\left(r^{4}\left(3-15x^{2}\right)+6r^{3}\left(5x^{3}+x\right)+r^{2}\left(1-39x^{2}\right)+16rx-2\right)\right]\\
A_{44}^{\mathrm{x}} & = & \frac{1}{32\left(r^{2}-2rx+1\right)}\left[xW_{p}\left(r^{5}\left(231x^{6}-315x^{4}+105x^{2}-5\right)+r^{4}\left(-462x^{7}+252x^{5}+210x^{3}-80x\right)\right.\right.\nonumber \\
 &  & +r^{3}\left(987x^{6}-945x^{4}+105x^{2}+13\right)-2r^{2}x\left(399x^{4}-370x^{2}+51\right)+2r\left(145x^{4}-114x^{2}+9\right)\nonumber \\
 &  & \left.+8x\left(3-5x^{2}\right)\right)+rW_{*}\left(r^{4}\left(-231x^{7}+315x^{5}-105x^{3}+5x\right)+r^{3}\left(609x^{6}-735x^{4}+195x^{2}-5\right)\right.\nonumber \\
 &  & \left.-12r^{2}x\left(49x^{4}-50x^{2}+9\right)+2r\left(125x^{4}-102x^{2}+9\right)+8x\left(3-5x^{2}\right)\right)\nonumber \\
 &  & +2\left(r^{4}\left(35x^{4}-30x^{2}+3\right)+r^{3}\left(-70x^{5}+20x^{3}+18x\right)\right.\nonumber \\
 &  & +r^{2}\left(115x^{4}-66x^{2}-1\right)+r\left(32x-64x^{3}\right)\nonumber \\
 &  & \left.\left.+12x^{2}-4\right)\right],
\end{eqnarray}
and
\begin{eqnarray}
B_{00}^{\mathrm{x}} & = & \frac{1}{42\left(r^{4}+r^{2}\left(2-4x^{2}\right)+1\right)}\left[21x^{2}W_{p}^{2}\left(-r^{4}+r^{2}\left(4x^{2}-2\right)-1\right)\right.\nonumber \\
 &  & +42\left(r^{2}+1\right)x^{2}W_{p}\left(r^{4}+r^{2}\left(2-4x^{2}\right)+1\right)+2r^{4}\left(28x^{4}-59x^{2}+10\right)+4r^{2}\left(38x^{4}-22x^{2}+5\right)\nonumber \\
 &  & -6rW_{*}\left(7r^{5}x^{2}+r^{4}x\left(10x^{2}-17\right)-2r^{3}\left(4x^{4}+8x^{2}-5\right)+2r^{2}x\left(9x^{2}-2\right)+r\left(10-3x^{2}\right)\right.\nonumber \\
 &  & \left.\left.-7x\right)-42x^{2}\right]\\
B_{01}^{\mathrm{x}} & = & \frac{x^{2}-1}{14\left(r^{2}-2rx+1\right)^{2}\left(r^{2}+2rx+1\right)}\left[7x^{2}W_{p}^{2}\left(r^{2}-2rx+1\right)^{2}\left(r^{2}+2rx+1\right)\right.\nonumber \\
 &  & +x^{2}W_{p}\left(-14r^{8}+28r^{7}x+r^{6}\left(44x^{2}-37\right)+r^{5}\left(46x-88x^{3}\right)+r^{4}\left(72x^{2}-51\right)\right.\nonumber \\
 &  & \left.-56r^{3}x\left(x^{2}-1\right)+7r^{2}\left(4x^{2}-5\right)+14rx-7\right)\nonumber \\
 &  & +rW_{*}\left(14r^{7}x^{2}+r^{6}x\left(2x^{2}-37\right)+r^{5}\left(-44x^{4}+21x^{2}+23\right)+r^{4}x\left(16x^{4}+110x^{2}-63\right)\right.\nonumber \\
 &  & \left.\left.+r^{3}\left(-44x^{4}-38x^{2}+40\right)+3r^{2}x\left(8x^{2}-15\right)+r\left(11x^{2}+17\right)-7x\right)\right]\\
B_{02}^{\mathrm{x}} & = & \frac{3\left(x^{2}-1\right)^{2}}{112\left(r^{2}-2rx+1\right)^{2}\left(r^{2}+2rx+1\right)}\left[-7x^{2}W_{p}^{2}\left(r^{2}-2rx+1\right)^{2}\left(r^{2}+2rx+1\right)\right.\nonumber \\
 &  & +2r^{2}x^{2}W_{p}\left(7r^{6}-14r^{5}x+r^{4}\left(9-16x^{2}\right)+4r^{3}x\left(8x^{2}-1\right)+r^{2}\left(9-16x^{2}\right)-14rx+7\right)\nonumber \\
 &  & -2r^{2}W_{*}\left(7r^{6}x^{2}+r^{5}\left(6x^{3}-20x\right)+r^{4}\left(-16x^{4}-4x^{2}+13\right)\right.\nonumber \\
 &  & \left.\left.+r^{3}\left(50x^{3}-22x\right)+r^{2}\left(20-27x^{2}\right)-14rx+7\right)\right]
\end{eqnarray}
\begin{eqnarray}
B_{11}^{\mathrm{x}} & = & \frac{1}{14\left(r^{2}-2rx+1\right)^{2}\left(r^{2}+2rx+1\right)}\left[-21x^{4}W_{p}^{2}\left(r^{2}-2rx+1\right)^{2}\left(r^{2}+2rx+1\right)+3x^{2}W_{p}\left(14r^{8}x^{2}\right.\right.\nonumber \\
 &  & -28r^{7}x^{3}+r^{6}\left(-44x^{4}+25x^{2}+19\right)+r^{5}\left(88x^{5}-22x^{3}-38x\right)+r^{4}\left(-72x^{4}+11x^{2}+33\right)\nonumber \\
 &  & \left.+28r^{3}x\left(2x^{4}-1\right)+7r^{2}\left(-4x^{4}+x^{2}+3\right)-14r\left(x^{3}+x\right)+7\left(x^{2}+1\right)\right)\nonumber \\
 &  & +2\left(r^{6}\left(28x^{4}-59x^{2}+10\right)-2r^{5}x\left(28x^{4}-59x^{2}+10\right)+r^{4}\left(104x^{4}-103x^{2}+20\right)\right.\nonumber \\
 &  & \left.-4r^{3}x\left(38x^{4}-22x^{2}+5\right)+r^{2}\left(76x^{4}-65x^{2}+10\right)+42rx^{3}-21x^{2}\right)\nonumber \\
 &  & -3rW_{*}\left(14r^{7}x^{4}+r^{6}x\left(2x^{4}-47x^{2}+3\right)+r^{5}\left(-44x^{6}+9x^{4}+52x^{2}-3\right)\right.\nonumber \\
 &  & +r^{4}x\left(16x^{6}+126x^{4}-53x^{2}-19\right)+r^{3}\left(-44x^{6}-82x^{4}+56x^{2}\right)\nonumber \\
 &  & \left.\left.+r^{2}x\left(24x^{4}-21x^{2}-17\right)+r\left(11x^{4}+28x^{2}+3\right)-7\left(x^{3}+x\right)\right)\right]\\
B_{12}^{\mathrm{x}} & = & -\frac{x^{2}-1}{56\left(r^{2}-2rx+1\right)^{2}\left(r^{2}+2rx+1\right)}\left[-7\left(15x^{2}+1\right)x^{2}W_{p}^{2}\left(r^{2}-2rx+1\right)^{2}\left(r^{2}+2rx+1\right)\right.\nonumber \\
 &  & +2x^{2}W_{p}\left(21r^{8}\left(5x^{2}+3\right)-42r^{7}x\left(5x^{2}+3\right)+r^{6}\left(-240x^{4}-225x^{2}+353\right)\right.\nonumber \\
 &  & +20r^{5}x\left(24x^{4}+33x^{2}-29\right)+r^{4}\left(-240x^{4}-497x^{2}+513\right)+r^{3}\left(334x^{3}-446x\right)\nonumber \\
 &  & \left.+r^{2}\left(279-167x^{2}\right)-112rx+56\right)+4\left(r^{6}\left(12x^{2}-19\right)+r^{5}\left(38x-24x^{3}\right)\right.\nonumber \\
 &  & \left.+r^{4}\left(52x^{2}-45\right)+r^{3}\left(52x-80x^{3}\right)+r^{2}\left(40x^{2}-33\right)+14rx-7\right)\nonumber \\
 &  & -2rW_{*}\left(21r^{7}x^{2}\left(5x^{2}+3\right)+6r^{6}x\left(15x^{4}-67x^{2}-11\right)+r^{5}\left(-240x^{6}-420x^{4}+587x^{2}+3\right)\right.\nonumber \\
 &  & +r^{4}\left(942x^{5}+84x^{3}-326x\right)+r^{3}\left(-869x^{4}+469x^{2}+36\right)-2r^{2}x\left(5x^{2}+128\right)\nonumber \\
 &  & \left.\left.+r\left(233x^{2}+33\right)-56x\right)\right]\\
B_{13}^{\mathrm{x}} & = & \frac{3}{16}\left(x^{2}-1\right)^{2}\left[10r^{2}x^{2}W_{p}-x^{2}W_{p}^{2}-\frac{2r^{2}W_{*}\left(5r^{2}x^{2}-7rx+2\right)}{r^{2}-2rx+1}\right]\\
B_{22}^{\mathrm{x}} & = & \frac{1}{112\left(r^{2}-2rx+1\right)^{2}\left(r^{2}+2rx+1\right)}\left[-7\left(35x^{4}-6x^{2}-5\right)x^{2}W_{p}^{2}\left(r^{2}-2rx+1\right)^{2}\left(r^{2}+2rx+1\right)\right.\nonumber \\
 &  & +2x^{2}W_{p}\left(7r^{8}\left(35x^{4}+10x^{2}-21\right)-14r^{7}x\left(35x^{4}+10x^{2}-21\right)\right.\nonumber \\
 &  & -r^{6}\left(560x^{6}+565x^{4}-1802x^{2}+677\right)+4r^{5}x\left(280x^{6}+405x^{4}-866x^{2}+265\right)\nonumber \\
 &  & -r^{4}\left(560x^{6}+1557x^{4}-2666x^{2}+885\right)+2r^{3}x\left(747x^{4}-934x^{2}+355\right)\nonumber \\
 &  & \left.-3r^{2}\left(249x^{4}-386x^{2}+137\right)-112rx\left(4x^{2}-1\right)+56\left(4x^{2}-1\right)\right)\nonumber \\
 &  & +8\left(r^{6}\left(36x^{4}-97x^{2}+19\right)-2r^{5}x\left(36x^{4}-97x^{2}+19\right)+r^{4}\left(220x^{4}-223x^{2}+45\right)\right.\nonumber \\
 &  & \left.-4r^{3}x\left(92x^{4}-63x^{2}+13\right)+r^{2}\left(184x^{4}-175x^{2}+33\right)+14rx\left(7x^{2}-1\right)-49x^{2}+7\right)\nonumber \\
 &  & -2rW_{*}\left(7r^{7}x^{2}\left(35x^{4}+10x^{2}-21\right)+2r^{6}x\left(105x^{6}-580x^{4}+127x^{2}+96\right)\right.\nonumber \\
 &  & -r^{5}\left(560x^{8}+1020x^{6}-2647x^{4}+854x^{2}+45\right)+2r^{4}x\left(1195x^{6}-147x^{4}-959x^{2}+331\right)\nonumber \\
 &  & -r^{3}\left(2577x^{6}-2038x^{4}+169x^{2}+132\right)+r^{2}\left(294x^{5}-904x^{3}+442x\right)\nonumber \\
 &  & \left.\left.+r\left(677x^{4}-86x^{2}-87\right)+56x\left(1-4x^{2}\right)\right)\right]\\
B_{23}^{\mathrm{x}} & = & \frac{x^{2}-1}{8\left(r^{2}-2rx+1\right)}\left[3\left(5x^{2}-1\right)x^{2}W_{p}^{2}\left(r^{2}-2rx+1\right)+2x^{2}W_{p}\left(-5r^{4}\left(7x^{2}-3\right)+10r^{3}x\left(7x^{2}-3\right)\right.\right.\nonumber \\
 &  & \left.+r^{2}\left(9-35x^{2}\right)+12rx-6\right)+4\left(r^{2}-2rx+1\right)+2rW_{*}\left(5r^{3}x^{2}\left(7x^{2}-3\right)+r^{2}\left(21x-65x^{3}\right)\right.\nonumber \\
 &  & \left.\left.+6r\left(6x^{2}-1\right)-6x\right)\right]\\
B_{33}^{\mathrm{x}} & = & -\frac{1}{16\left(r^{2}-2rx+1\right)}\left[\left(35x^{4}-30x^{2}+3\right)x^{2}W_{p}^{2}\left(r^{2}-2rx+1\right)+2x^{2}W_{p}\left(r^{4}\left(-63x^{4}+70x^{2}-15\right)\right.\right.\nonumber \\
 &  & \left.+2r^{3}x\left(63x^{4}-70x^{2}+15\right)+r^{2}\left(-63x^{4}+50x^{2}-3\right)+8rx\left(5x^{2}-3\right)-20x^{2}+12\right)\nonumber \\
 &  & +8\left(3x^{2}-1\right)\left(r^{2}-2rx+1\right)+2rW_{*}\left(r^{3}x^{2}\left(63x^{4}-70x^{2}+15\right)+r^{2}\left(-133x^{5}+130x^{3}-21x\right)\right.\nonumber \\
 &  & \left.\left.+r\left(90x^{4}-72x^{2}+6\right)+4x\left(3-5x^{2}\right)\right)\right].
\end{eqnarray}

\section{Impact of Different Corrections}

In this section, we analyze the impact of various corrections on the best-fit estimation of the linear growth rate. As an example, we consider the counter-term type in Eq.~(\ref{eq:ct1}) with a smoothing scale of $R_{\mathrm{s}} = 15h^{-1} \mathrm{Mpc}$. We examine the effects of grid correction, variations in $k_{\mathrm{min}}$, and binning effect correction.

\begin{figure}
\begin{center}
\includegraphics[width=8.5cm]{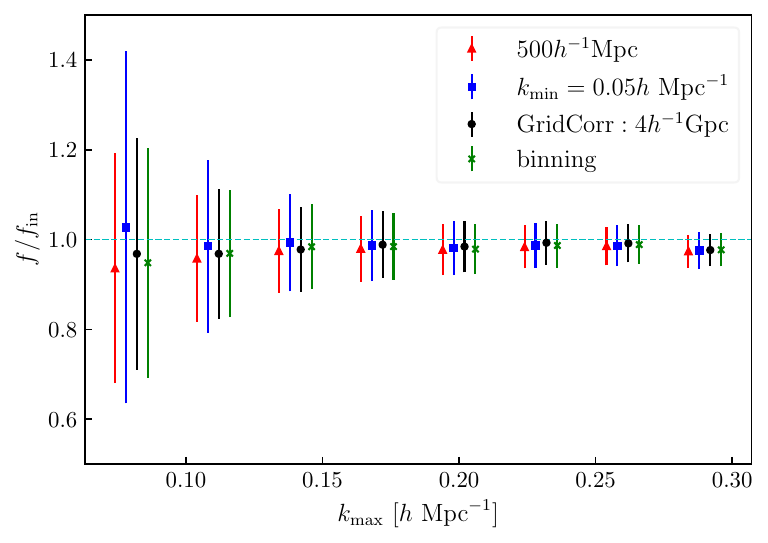}
\caption{\label{fig_bestfit_test} Best-fit values of the growth rate under different correction tests for the counter-term type in Eq.~(\ref{eq:ct1}). Red points represent the uncorrected results. Blue points show the results for $k_{\mathrm{min}} = 0.05h \mathrm{Mpc}^{-1}$ (compared to $k_{\mathrm{min}} = 0.01h \mathrm{Mpc}^{-1}$ for other cases). Black points represent the results after applying grid correction based on Eq.~(\ref{eq:grid-corr}), while green points correspond to the binning effect correction following Eq.~(\ref{eq:binning}).}
\end{center}
\end{figure}

Figure~\ref{fig_bestfit_test} presents the best-fit values of the linear growth rate obtained from these tests, along with the results from an $N$-body simulation with a box length of $L = 500h^{-1} \mathrm{Mpc}$ for comparison. The range of $k_{\mathrm{max}}$ is kept consistent with Figure~\ref{fig_bestfit}.

Since the measured data from the simulation with $L = 500h^{-1} \mathrm{Mpc}$ exhibits irregular fluctuations at large scales, we investigate the impact of removing several data points at large scales by varying $k_{\mathrm{min}}$. This approach helps to obtain more unbiased fitting results, albeit at the cost of increased statistical errors. The blue points in Figure~\ref{fig_bestfit_test} represent the results for $k_{\mathrm{min}} = 0.05h \mathrm{Mpc}^{-1}$, demonstrating a balance between fitting bias and statistical uncertainty.

Inspired by this observation, we apply a grid correction based on Eq.~(\ref{eq:grid-corr}) to the simulated data from the $500h^{-1} \mathrm{Mpc}$ box, incorporating data from a larger simulation box with $L = 4h^{-1} \mathrm{Gpc}$. The best-fit results after applying this grid correction are shown as black points in Figure~\ref{fig_bestfit_test}, revealing an improvement in the unbiasedness of the best-fit value of $f$ at large scales compared to the red points. However, noticeable deviations from the input value persist, particularly at large scales. To further refine the results, we introduce a binning correction following Eq.~(\ref{eq:binning}), ensuring better consistency between theoretical calculations and simulated data. The green points in Figure~\ref{fig_bestfit_test} illustrate that the best-fit values after binning correction are closer to the input value. Finally, after implementing both grid correction and binning correction, we achieve the most accurate best-fit results overall, as indicated by the black points in Figure~\ref{fig_bestfit}.

\bibliography{draft}{}
\bibliographystyle{aasjournal}

\end{document}